\documentclass{class}
\usepackage{graphicx,amsmath,bm,mathrsfs,xcolor}
\usepackage{epstopdf, epsfig,psfrag}
\usepackage{cases}
\usepackage[normalem]{ulem}

\newcommand{\rd}{\mathrm{d}}
\newcommand{\req}{r_{\mathrm{eq}}}

\shorttitle{Droplets in isotropic turbulence: deformation and breakup statistics}
\shortauthor{S. S. Ray and D. Vincenzi}

\title{Droplets in isotropic turbulence: deformation and breakup statistics}

\author{
Samriddhi Sankar Ray\aff{1}
 \and 
Dario Vincenzi\aff{2}
  \corresp{\email{dario.vincenzi@unice.fr}}
}

\affiliation{
\aff{1}International Centre for Theoretical Sciences, Tata Institute of Fundamental
Research, Bangalore 560089, India
\aff{2}Universit\'e C\^ote d'Azur, CNRS, LJAD, 06108 Nice, France
}

\begin{document}

\maketitle

\begin{abstract}
The statistics of the 
deformation and breakup of neutrally buoyant sub-Kolmogorov ellipsoidal
drops is investigated via Lagrangian simulations of homogeneous
isotropic turbulence.
The mean lifetime of a drop is also studied as a function of the initial drop
size and the capillary number. A vector model of drop previously introduced by
Olbricht, Rallison \& Leal 
[\textit{J. Non-Newtonian Fluid Mech.} \textbf{10}, 291--318 (1982)]
is used to predict the behaviour of the above
quantities analytically.
\end{abstract}


\section{Introduction}

The dispersion of drops of one fluid in another fluid
that is turbulent and immiscible with the first
has numerous applications.
Emulsion processing in chemical engineering, for instance,
often uses turbulent
flow conditions~\citep{W93,SS03}, and
the design of efficient emulsion apparatuses
requires a detailed understanding of
single-drop dynamics in turbulent flows~\citep{Wetal05}.


The theory of \cite{K49} and \cite{H95} predicts two different regimes 
according to whether a drop is larger or smaller, respectively, than the 
Kolmogorov dissipation scale~$\ell_K$. In the former case,
the dynamics of the drop results from the competition 
between the inertial hydrodynamic stress, which distorts the drop,
and the stress due to surface tension, which restores
the drop to its equilibrium configuration. In the latter case,
the competition is between surface tension and the viscous stress.
The literature on drop dynamics in turbulent flows 
has largely focused on drops whose size lies in
the inertial range. The sub-Kolmogorov regime, albeit
difficult to examine both experimentally and numerically,
is of practical significance as well.
For viscous oils, turbulent emulsification is indeed known to be more
efficient in the sub-Kolmogorov regime~\citep{Vetal07}. In addition,
even if the initial drop sizes are larger than $\ell_K$,
in high-Reynolds-number flows
subsequent breakups can generate sub-Kolmogorov drops
at long times \citep{CBLC03}.
Another
mechanism for the formation of small drops in a turbulent flow
was recently reported in \citet{PWKPB17}: it consists in the
nucleation of microdroplets
in the wake of a large cold drop crossing a supersaturated environment. 
Drops smaller than $\ell_K$
were also used as tracers in laboratory experiments
with the purpose of examining the statistics of the Lagrangian acceleration
in turbulent flows~\citep{AGW08,ACW08}.

The deformation and breakup of a drop in a chaotic flow were first studied
by \citet{TO91} and \citet{MTO91} by means of
a `journal-bearing' flow generated
by the periodic motion of two rotating eccentric cylinders.
The fluid trajectories are chaotic in this flow, so a
drop becomes highly stretched,
folds and eventually breaks. Subsequent breakups of the drop
fragments lead to a population of
drops with different sizes; various modes of breakup were observed,
including capillary-wave instabilities, necking, end- and fold-pinching.

\cite{CBLC03} studied the dynamics
of a sub-Kolmogorov drop in a numerical
simulation of homogeneous and isotropic turbulence
at moderate Reynolds number.
The trajectory of the
centre of mass of the drop was approximated by a fluid trajectory, under
the assumption of a small density
contrast between the fluids inside and outside the drop.
The dynamics of the drop was calculated via a boundary integral approach 
by using the Stokes equations with appropriate
boundary conditions at the drop interface
and with a far field given by a linear 
expansion of the external turbulent flow about the position of the 
centre of mass. The statistics of drop length, orientation, and
breakup was studied as a function of the 
viscosity ratio between the inner and outer fluids
and of the capillary number. This latter determines
the relative intensity of the viscous and surface-tension forces.
It was shown, in particular, that under moderate-deformation conditions
drop reorientation is mainly due to the deformation of the
drop surface rather than the rotation of the drop by the flow.

For high Reynolds numbers,
the direct numerical simulation of sub-Kolmogorov
drops is still impractical with the 
available computational facilities, especially when 
a very large number of drops needs to be considered in order to
resolve the statistical properties of drop dynamics.
An alternative approach consists in using simplified models of drops.
\citet{BMV14} coupled the model of \citet{MM98}, which describes 
neutrally buoyant
ellipsoidal drops, with a Lagrangian simulation of high-Reynolds-number,
homogeneous and isotropic turbulence. The model of \citet{MM98}
was originally derived for linear flows but can be applied to turbulent
flows if the Reynolds number at the scale of the drop is smaller than unity, i.e.
the size of the drop is smaller than $\ell_K$.
This approach allowed the authors
to obtain a detailed statistical characterization of drop deformation and
orientation. In particular,
the statistics of the deformation was related
to that of the stretching rates of the flow via
an analogy between the model of \citet{MM98} and the Oldroyd-B model
for flexible polymers \citep[e.g.][]{BHAC87}. A critical
capillary number for breakup
was thus identified for the case in which the
viscosities of the fluids inside and outside the drop coincide.
\citet{SLV16} recently applied 
the model of \citet{MM98}
to a turbulent Taylor--Couette flow, in order to examine
the dependence of drop dynamics on the
flow geometry.

The goal of the present study is to further investigate and elucidate
the statistical properties of drop deformation and breakup
in the sub-Kolmogorov regime. To this end,
we follow the approach proposed by \citet*{BMV14} and use the
model of \cite{MM98} in combination with Lagrangian simulations
of homogeneous and isotropic turbulence.
We perform a detailed numerical analysis 
of the time-dependent and time-integrated probability density functions of drop size 
as a function of the capillary number, the viscosity ratio between the inner 
and outer fluids, and the initial drop-size distribution. 
We also study the breakup rate and the mean lifetime of a drop as a 
function of the same quantities.
The results of the numerical simulations are then 
derived analytically by means of a vector model of drop originally proposed 
by \citet{ORL82}.

\section{Deformation and breakup statistics}

The model of \citet{MM98} assumes that
both the fluid of which the drop is composed and the fluid in which it is 
immersed are Newtonian. 
The drop is neutrally buoyant and is transported passively (i.e., it does
not affect the surrounding flow),
it is ellipsoidal at all times,
and its volume is preserved. 
In addition, the flow about the drop is incompressible and linear. 
This latter assumption is appropriate
for turbulent flows if the size of the drop is smaller than $\ell_K$.
The volume fraction is very low, so that hydrodynamic interactions between 
drops are negliglible and attention can be directed to the dynamics of a
single drop.

The shape and the orientation
of the drop are described by a
second-rank symmetric positive-definite tensor $\mathsfbi{M}$, whose
eigenvectors are the semi-axes of the drop and whose eigenvalues
$m_1^2\geqslant m_2^2 \geqslant m_3^2$
yield the squared lengths of the same semi-axes.
The centre of mass of the drop evolves as a tracer, while
the Lagrangian evolution of $\mathsfbi{M}$ is given by the following equation:
\begin{equation}
\label{eq:M}
\dot{\mathsfbi{M}}=\mathsfbi{G}\mathsfbi{M}+\mathsfbi{M}\mathsfbi{G}^\top
-\dfrac{f_1(\mu)}{\tau}[\mathsfbi{M}-g(\mathsfbi{M})\mathsfbi{I}],
\end{equation}
where $\mathsfbi{G}=f_2(\mu)\mathsfbi{S}
+\mathsfbi{\Omega}$ is an effective velocity gradient;
$\mathsfbi{\Omega}=[\bnabla\bm u-(\bnabla\bm u)^\top]/2$
and $\mathsfbi{S}=[\bnabla\bm u+(\bnabla\bm u)^\top]/2$  are
the vorticity and
rate-of-strain tensors evaluated at the centre of mass of the drop, 
respectively. Note that here $(\bnabla\bm u)_{ij}=\partial_j u_i$.
The coefficients $f_1(\mu)$ and $f_2(\mu)$ depend on the ratio $\mu$ of the 
viscosity of the drop and that of the external fluid and were chosen in such a
way as to match theoretical predictions for small capillary numbers
\citep{MM98}: 
\begin{equation}
\label{eq:fs}
f_1(\mu)=\dfrac{40(\mu+1)}{(2\mu+3)(19\mu+16)}, 
\quad 
f_2(\mu)=\dfrac{5}{2\mu+3}.
\end{equation}
Note that $f_2(1)=1$ and hence, for $\mu=1$, $\mathsfbi{G}=\bnabla\bm u$.
The last term in \eqref{eq:M} describes the capillary
relaxation to the spherical shape
with a time scale $\tau$. Thanks to an appropriate choice of the function
$g(\mathsfbi{M})$, the same term
enforces that $\det\mathsfbi{M}$ is constant in time and hence
the volume of the drop is preserved.
The function $g(\mathsfbi{M})$ 
has the form $g(\mathsfbi{M})=3\mathrm{III}_M/\mathrm{II}_M$, where 
$\mathrm{II}_M$ and $\mathrm{III}_M$ are the second and third invariants 
of $\mathsfbi{M}$,
i.e. $\mathrm{II}_M=[(\operatorname{tr}\mathsfbi{M})^2-\operatorname{tr}\mathsfbi{M}^2]/2$
and $\mathrm{III}_M=\det\mathsfbi{M}$.
\citet{MM98} also proposed an improved expression of $f_2$ that depends
on the capillary number and more accurately describes the deformations observed in experiments for
large strains and high viscosity ratios. For the sake of simplicity, here we
use the coefficients given in \eqref{eq:fs}; the improved version of the
model of \cite{MM98} is discussed in \S~\ref{sect:improved-f2}.

This Section provides insight into the statistics of drop
deformation and breakup in three-dimensional homogeneous isotropic
turbulence. We obtain such a turbulent flow by 
performing direct numerical simulations of the three-dimensional 
Navier--Stokes equation 
\begin{equation}
\label{eq:NS}
\partial_t\bm u+\bm u\bcdot\bnabla\bm u=-\bnabla p+\nu\Delta\bm u+\bm F
\end{equation}
for the velocity field $\bm u$ (and pressure $p$) augmented with the incompressibility condition $\bnabla\bcdot\bm u = 0$. We use the 
standard, fully de-aliased pseudo-spectral method on a cubic domain 
of size $2\upi$ with $512^3$  
collocation points and periodic boundary conditions. 
By using these boundary conditions, we do not take into account the interaction of the drops with 
the walls that confine the fluid.
The flow 
is driven to a non-equilibrium steady state by an external force $\bm F$ with a fixed 
energy input $\epsilon$. Our choice of $\epsilon$ and kinematic viscosity $\nu$ ensures 
a Taylor-scale Reynolds number ${\rm Re}_\lambda \approx 111$. 

In order to study the deformation of droplets in a turbulent flow, we seed our turbulent, statistically steady, 
flow with Lagrangian tracers and follow their trajectories, by using a trilinear-interpolation scheme 
to obtain the tracer velocity from the Eulerian velocity evaluated from Eq.~\eqref{eq:NS};
such trajectories define the motion of the center mass of the droplets. 
We refer the reader to \citet*{MJ17} for a more detailed description of our numerical procedure.

The capillary number is defined as $\mathit{Ca}=\lambda\tau$,
where $\lambda$ is the Lyapunov exponent of the flow. This latter represents
the average stretching rate in a turbulent flow and provides a measure
of the ability of the flow to deform a drop.
We calculate $\lambda$ by using the fluid velocity gradients along the trajectories and
obtain $\lambda\approx 4.22\approx 0.15\tau_\eta^{-1}$ 
(where $\tau_\eta$ is the Kolmogorov 
time-scale associated with the flow), consistent with earlier 
results \citep{BBBCLMT06}. 
(Note that \citet*{BMV14} defined the capillary number in terms of
the root mean square of $\partial_x u_x$ instead of $\lambda$. However,
this fact only leads to a different proportionality factor in the definition of $\mathit{Ca}$,
since in isotropic turbulence $\sqrt{\langle(\partial_x u_x)^2\rangle}=1/\sqrt{15}\tau_\eta$ and hence
in our case $\sqrt{\langle(\partial_x u_x)^2\rangle}= 1.72 \lambda$.)

Equation \eqref{eq:M} is integrated by using the second-order
Adam--Bashforth method with same time step as for the Navier--Stokes equation.
The integration of~\eqref{eq:M} must preserve the positive-definite character
of $\mathsfbi{M}$. This is achieved by adapting to \eqref{eq:M}
the Cholesky-decomposition method proposed by \citet{VC03} 
(see the appendix for the details).
Unless otherwise stated, the initial condition is $\mathsfbi{M}(0)=
\mathsfbi{I}$.
As in \citet*{BMV14}, it is assumed that drops break
when their aspect ratio $\vert m_1/ m_3\vert$ exceeds 
a threshold value $\alpha$.
In view of the fact that 
we are only interested in the dynamics up to the first
breakup and do not consider secondary breakup events, 
drops are removed from the flow as soon as they break.
In the simulations presented below, 
the initial number of drops $N(0) =  10^6$.

\begin{figure}
\centering
\psfrag{y}[c]{$\mathscr{P}(m_1^2)$}
\psfrag{x}[c]{$m_1^2$}
\psfrag{a}[lb]{$(m_1^2)^{-1}$}
\includegraphics[width=0.50\textwidth]{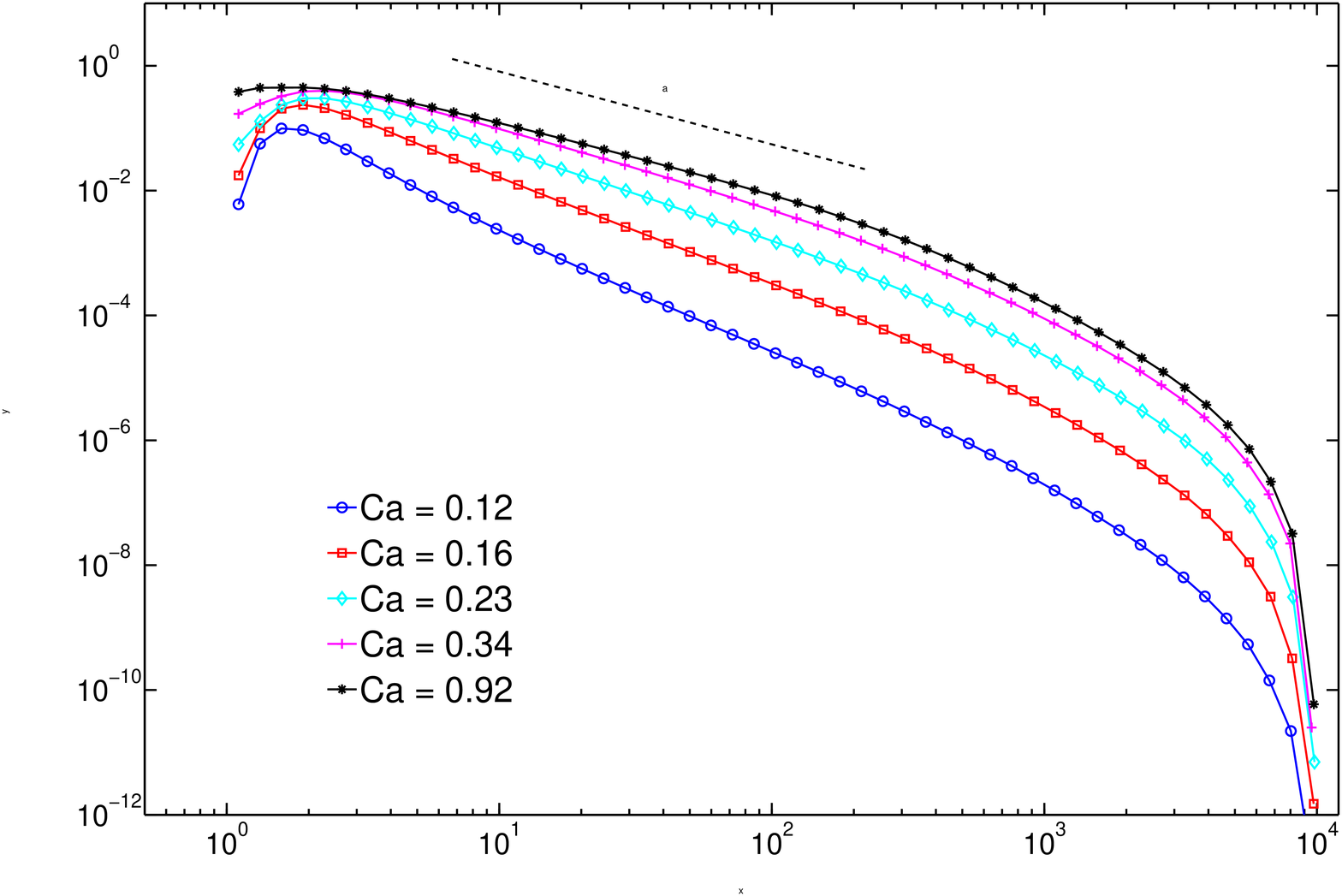}%
\hspace*{0.2cm}
\psfrag{y}[c]{$\mathscr{P}(m_1^2/m_3^2)$}
\psfrag{x}[c]{$m_1^2/m_3^2$}
\psfrag{a}[b]{$(m_1^2)^{-1}$}
\includegraphics[width=0.50\textwidth]{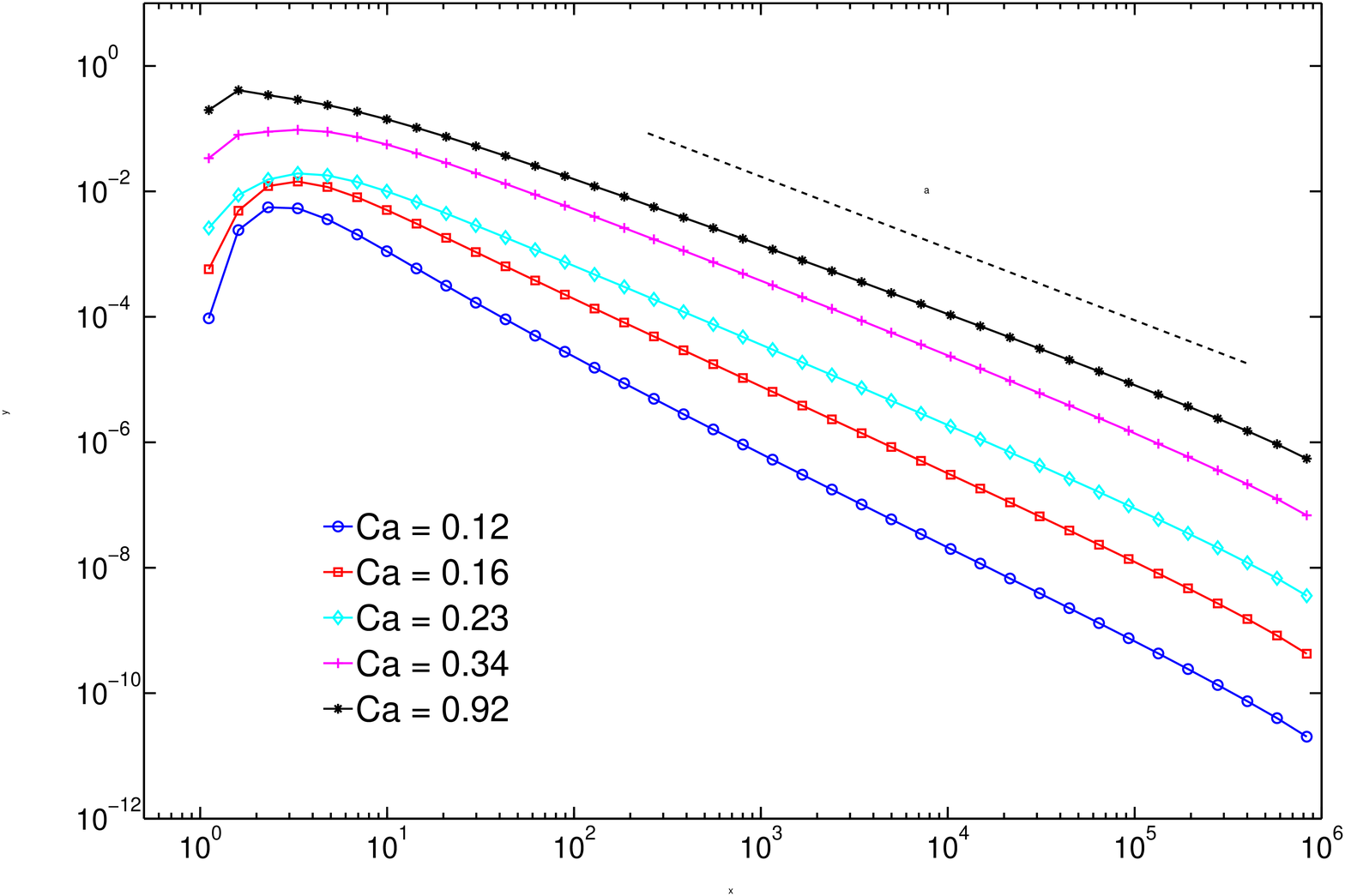}%
\caption{Time-integrated p.d.f. of (\textit{a}) the largest eigenvalue of $\mathsfbi{M}$
and (\textit{b}) the ratio of the largest and the smallest eigenvalue of $\mathsfbi{M}$
for $\mu=1$, $\alpha=10^3$, and different values of $\mathit{Ca}$. 
For this value of $\mu$, $Ca_c=0.23$ (see \S~\ref{sect:analytical}).
The p.d.f.s are artificially translated
in order to render their power-law behaviours more easily visible.}
\label{fig:m_1}
\end{figure}

The deformation of a drop is described
in terms of the statistics of $m_1^2$,
i.e. the squared length of the semi-major axis.
Let $p(m_1^2,t)$ be the p.d.f. of $m_1^2$ and 
$\mathscr{P}(m_1^2)\equiv\int_0^\infty p(m_1^2,t)\mathrm{d}t$
its time integral.
\citet*{BMV14} showed that
$\mathscr{P}(m_1^2)$ behaves as a power law for values of $m_1^2$ smaller
than its maximum value (it is easy to check that the conditions
$m_1^2\geqslant m_2^2\geqslant m_3^2$, $m_1^2 m_2^2 m_3^2=1$ and
$\vert m_1/m_3\vert\leqslant \alpha$ imply that $m_1^2$ is bounded and, more precisely,
$m_1^2\leqslant\alpha^{4/3}$).
The slope increases as a function of $\mathit{Ca}$ for small $\mathit{Ca}$
and saturates to $-1$ when 
$\mathit{Ca}$ exceeds a critical value $\mathit{Ca}_c$, 
which for $\mu=1$ was found to be $\mathit{Ca}_c=f_1(\mu)/2$.
Figure~\ref{fig:m_1}(\textit{a}) 
shows that the behaviour of $\mathscr{P}(m_1^2)$ is accurately reproduced in our simulations.
The power law is even clearer when the p.d.f of $m_1^2/m_3^2$ is considered (figure \ref{fig:m_1}(\textit{b})).

\begin{figure}
\psfrag{y}[c]{$N(t)/N(0)$}%
\psfrag{x}[c]{$t/\tau_\eta$}%
\psfrag{b}[c]{\scriptsize decay rate}%
\psfrag{a}[c]{\scriptsize $\mathit{Ca}/\mathit{Ca}_c$}%
\includegraphics[width=0.50\textwidth]{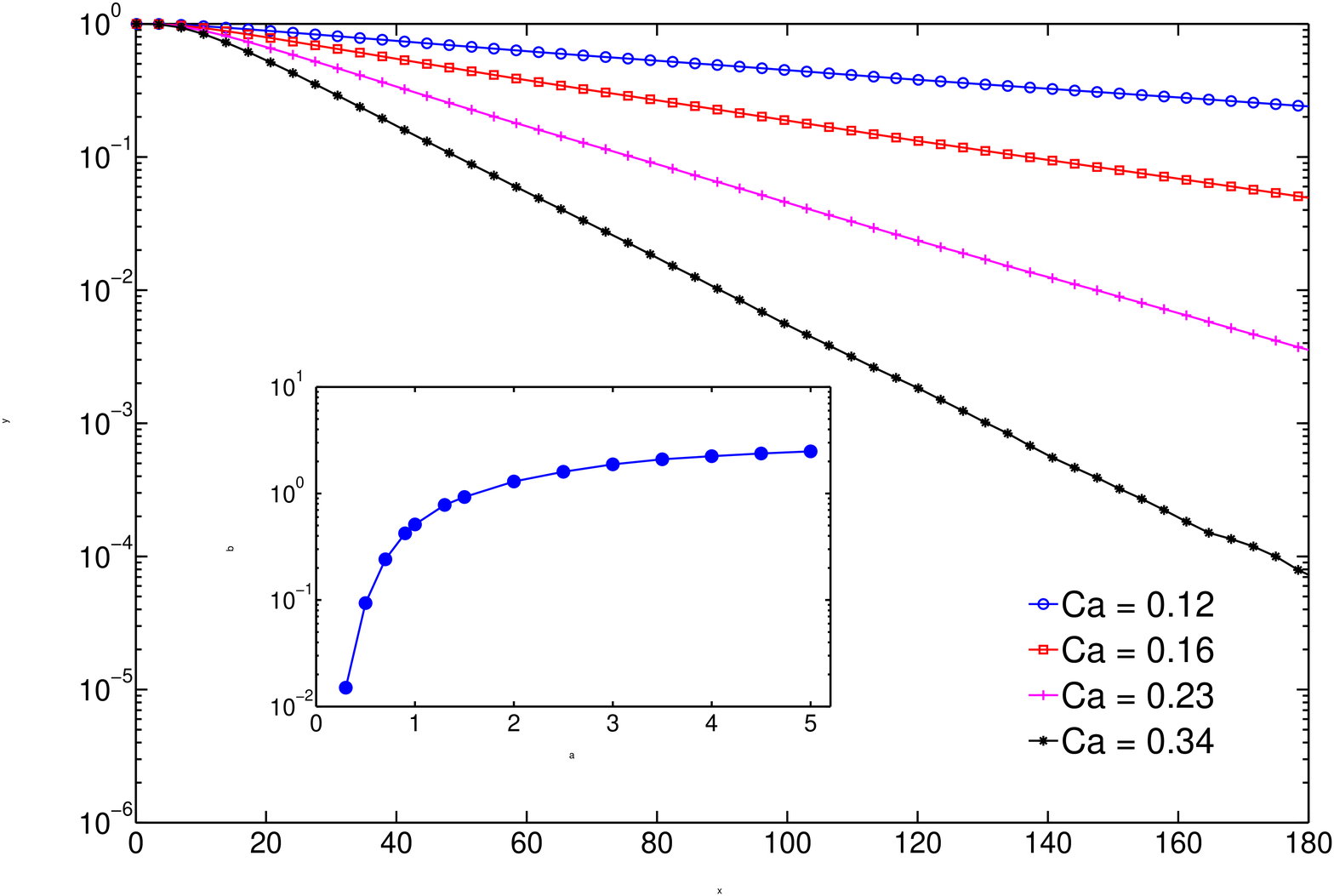}%
\hspace*{0.2cm}%
\psfrag{y}[c]{$p(m_1^2,t)$}
\psfrag{x}[c]{$m_1^2$}
\psfrag{a}[c]{$(m_1^2)^{-1}$}
\includegraphics[width=0.50\textwidth]{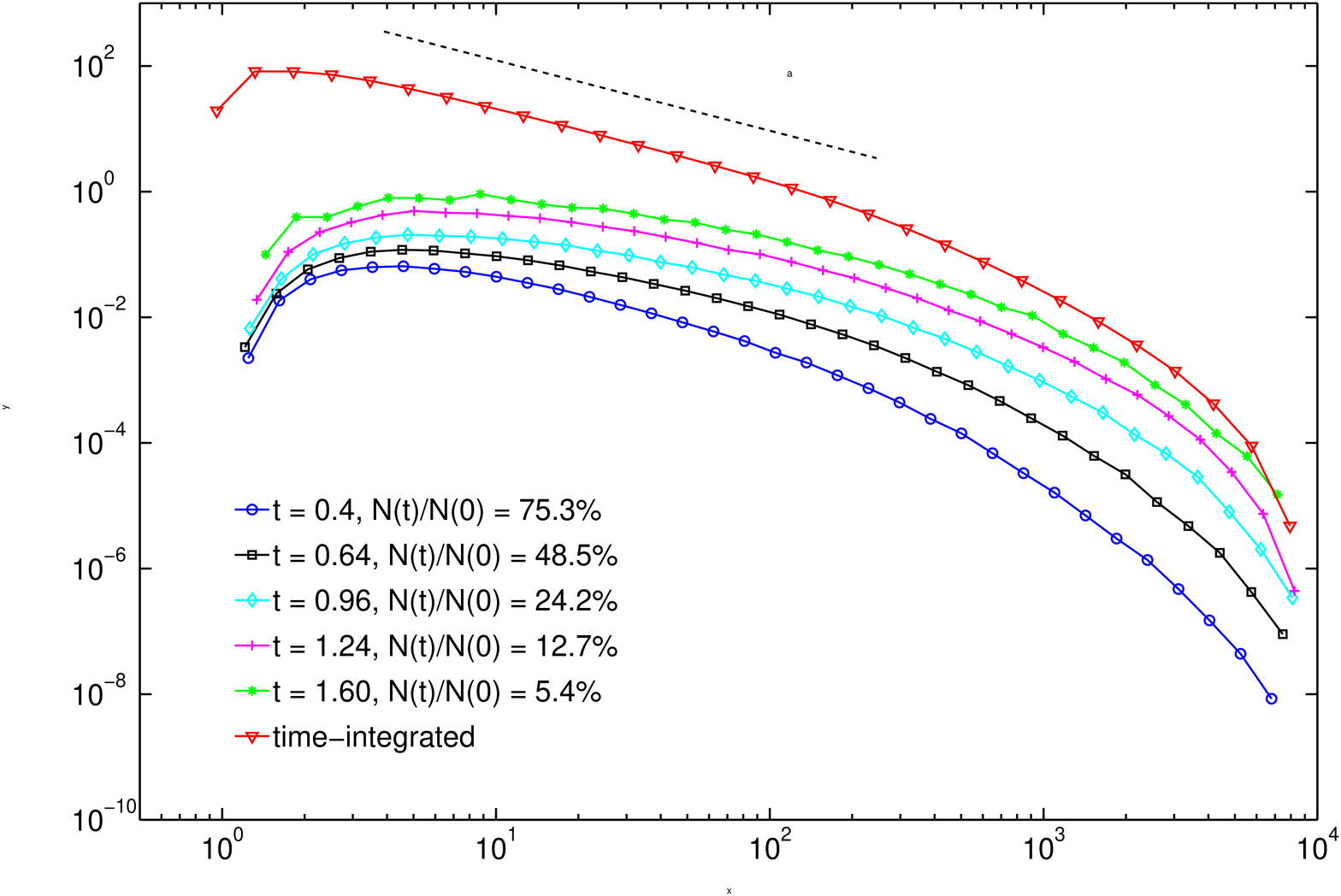}%
\caption{(\textit{a}) Fraction of surviving drops as a function of time for $\mu=1$, $\alpha=10^3$
and different values of $\mathit{Ca}$.  Time is rescaled by
the Kolmogorov time ($\tau_\eta$) of the flow. The inset shows the exponential
decay rate of the fraction of surviving drops vs the capillary number rescaled by its
critical value;
(\textit{b}) time-dependent p.d.f. of $m_1^2$ for $\mu=1$, $\alpha=10^3$,
$\mathit{Ca}=1$ and increasing time instants. 
In the legend, the fraction of drops surviving at time $t$ is also indicated.
The red curve is the time-integrated 
p.d.f. $\mathscr{P}(m_1^2)$ corresponding to the same parameters.
For the sake of comparison, the p.d.f.s are translated vertically.
}
\label{fig:time-dependence}
\end{figure}

It should be noted, however, that because of the breakups
the total number of drops decays in time.
The fraction of drops surviving at time $t$,
$N(t)/N(0)\equiv\int  p(m_1^2,t)\mathrm{d}m_1^2$,
indeeed decreases exponentially,
the decay rate growing rapidly
when $\mathit{Ca}$ exceeds $\mathit{Ca}_c$
(figure \ref{fig:time-dependence}(\textit{b})).
Accordingly, the statistics
of drop sizes is not stationary and
the p.d.f. of $m_1^2$ varies in time
(see figure~\ref{fig:time-dependence}(\textit{a}),
where the p.d.f.s are translated vertically in order to 
facilitate the comparison at different times).
At long times $p(m_1^2,t)$ reaches
an asymptotic shape, but this does not show any definite power-law 
behaviour.
The power law observed by \citet*{BMV14} is thus recovered only
when the time-integrated p.d.f is considered; indeed
the distributions shown in \citet*{BMV14} were obtained by averaging
over both the Lagrangian trajectories and time. 

\begin{figure}
\centering
\psfrag{y}[c]{$\mathscr{P}(m_1^2)$}
\psfrag{x}[c]{$m_1^2$}
\psfrag{a}[b]{$(m_1^2)^{-1}$}
\includegraphics[width=0.5\textwidth]{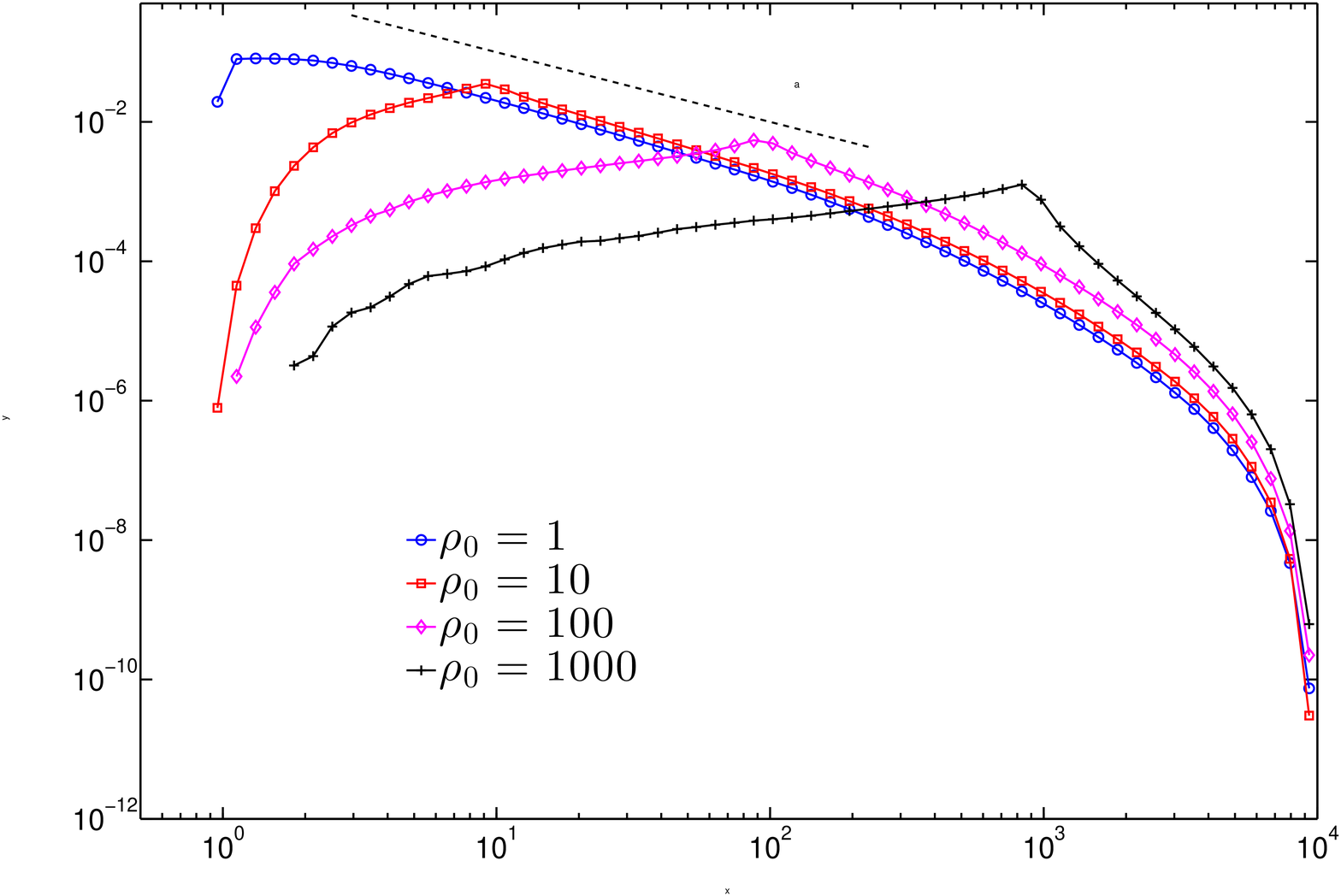}%
\hspace*{0.2cm}
\includegraphics[width=0.5\textwidth]{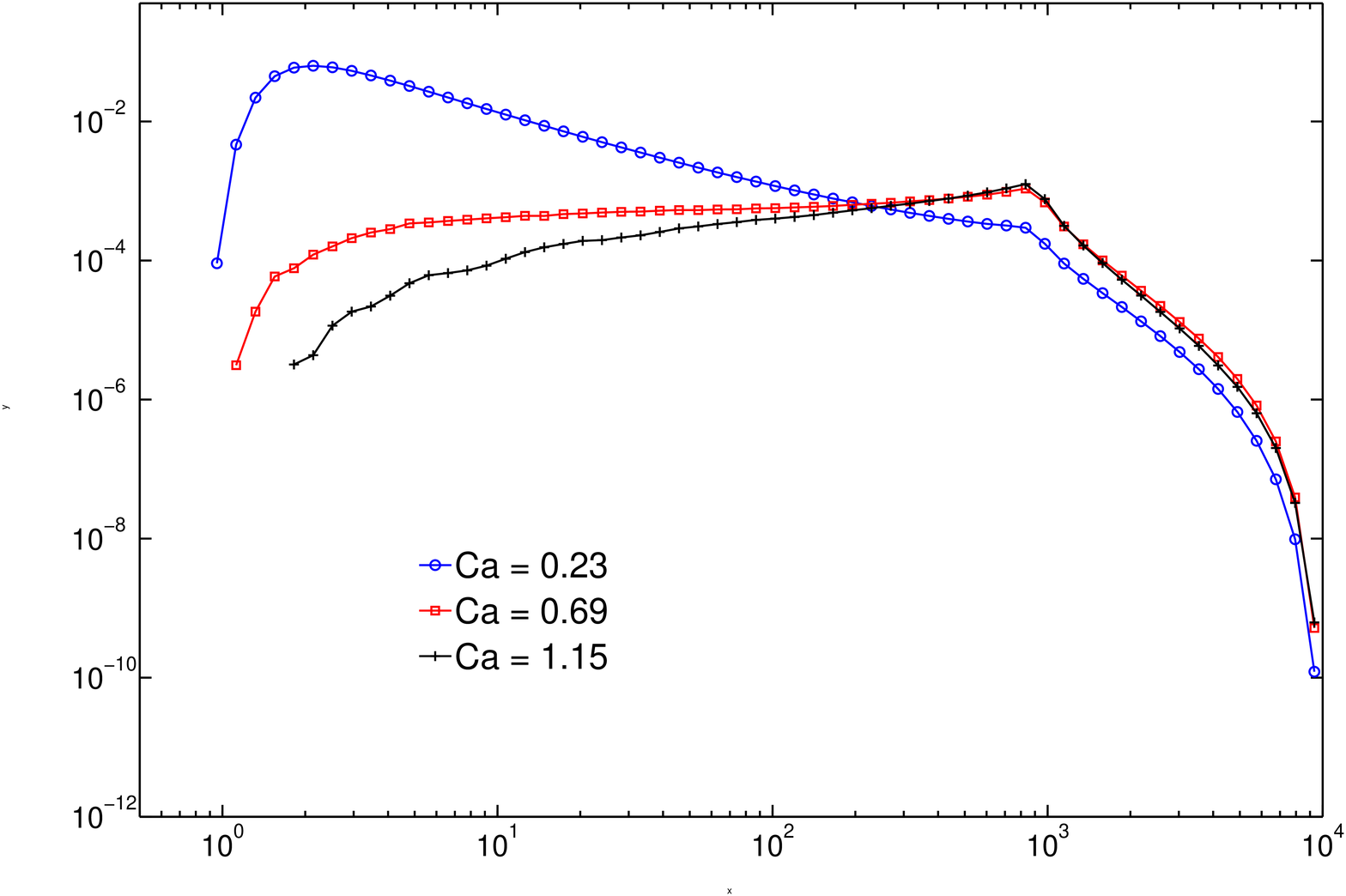}
\caption{(\textit{a}) Time-integrated p.d.f. of the largest eigenvalue of $\mathsfbi{M}$
for $\mu=1$, $\alpha=10^3$, $\mathit{Ca}=1$ and different values of $\rho_0$. 
The p.d.f.s are normalized to 1 to facilitate the comparison.
The dashed line is proportional to $(m_1^2)^{-1}$;
(\textit{b}) time-integrated p.d.f. of the largest eigenvalue of
$\mathsfbi{M}$ for $\mu=1$, $\rho_0=10^3$ and different values of $\mathit{Ca}$.
}
\label{fig:initial_condition}
\end{figure}

Since the dynamics of drops is not statistically stationary, $\mathscr{P}(m_1^2)$ may
depend on the initial shape of drops, namely on the value of the aspect ratio
at time $t=0$. We thus performed simulations in which the initial shape tensor
is $\mathsfbi{M}(0)=\operatorname{diag}(\rho_0,1,\rho_0^{-1})$, 
where $\rho_0>1$ is both the aspect ratio and the largest eigenvalue
of $\mathsfbi{M}$ at $t=0$. 
Two different behaviours are observed
depending on the value of $\mathit{Ca}$.
For small $\mathit{Ca}$, the shape of $\mathscr{P}(m_1^2)$ is not affected significantly
by the value of $\rho_0$ (not shown).
By contrast, for large $\mathit{Ca}$, the interval
over which $\mathscr{P}(m_1^2)\sim (m_1^2)^{-1}$ shrinks 
as $\rho_0$ is increased and
the drop volume is kept constant. In this case, indeed,
$\mathscr{P}(m_1^2)\sim (m_1^2)^{-1}$ only for $m_1^2\gg\rho_0$
(figure~\ref{fig:initial_condition}(\textit{a})). 
The $(m_1^2)^{-1}$ behaviour may therefore be difficult
to detect when $\rho_0$ approaches the critical aspect ratio for breakup.
In fact, when $\rho_0$ is sufficiently large a second power law emerges for $m_1^2\ll\rho_0$
whose slope increases as a function of $\mathit{Ca}$ and can turn from negative
to positive at large $\mathit{Ca}$ (figure~\ref{fig:initial_condition}(\textit{b})).

\begin{figure}
\centering
\psfrag{y}[c]{$\mathscr{P}(m_1^2)$}
\psfrag{x}[c]{$m_1^2$}
\psfrag{a}[lb]{$(m_1^2)^{-1}$}
\includegraphics[width=0.50\textwidth]{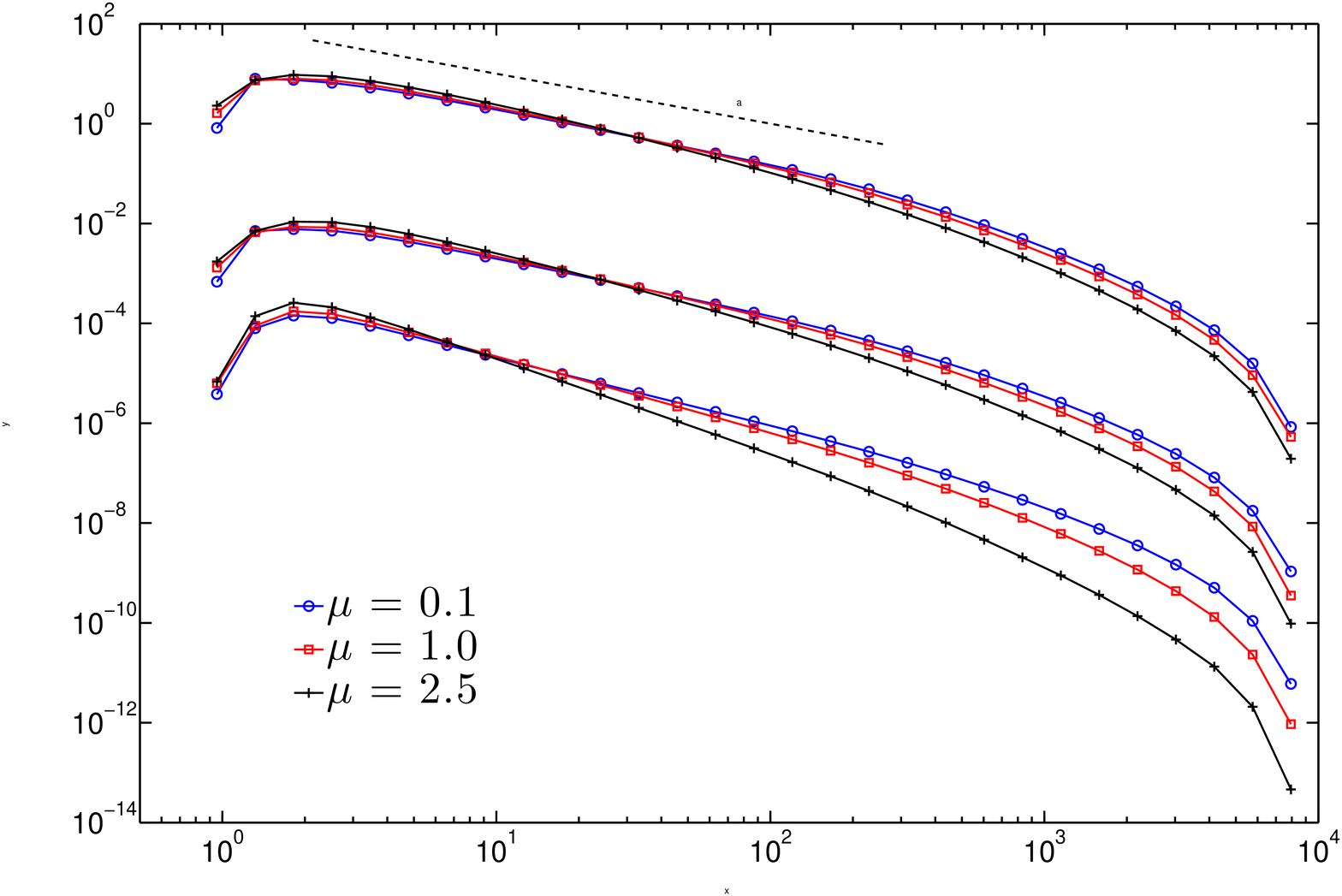}%
\hspace*{0.01cm}
\psfrag{y}[c]{decay rate}
\psfrag{x}[c]{$\mu$}
\includegraphics[width=0.5\textwidth]{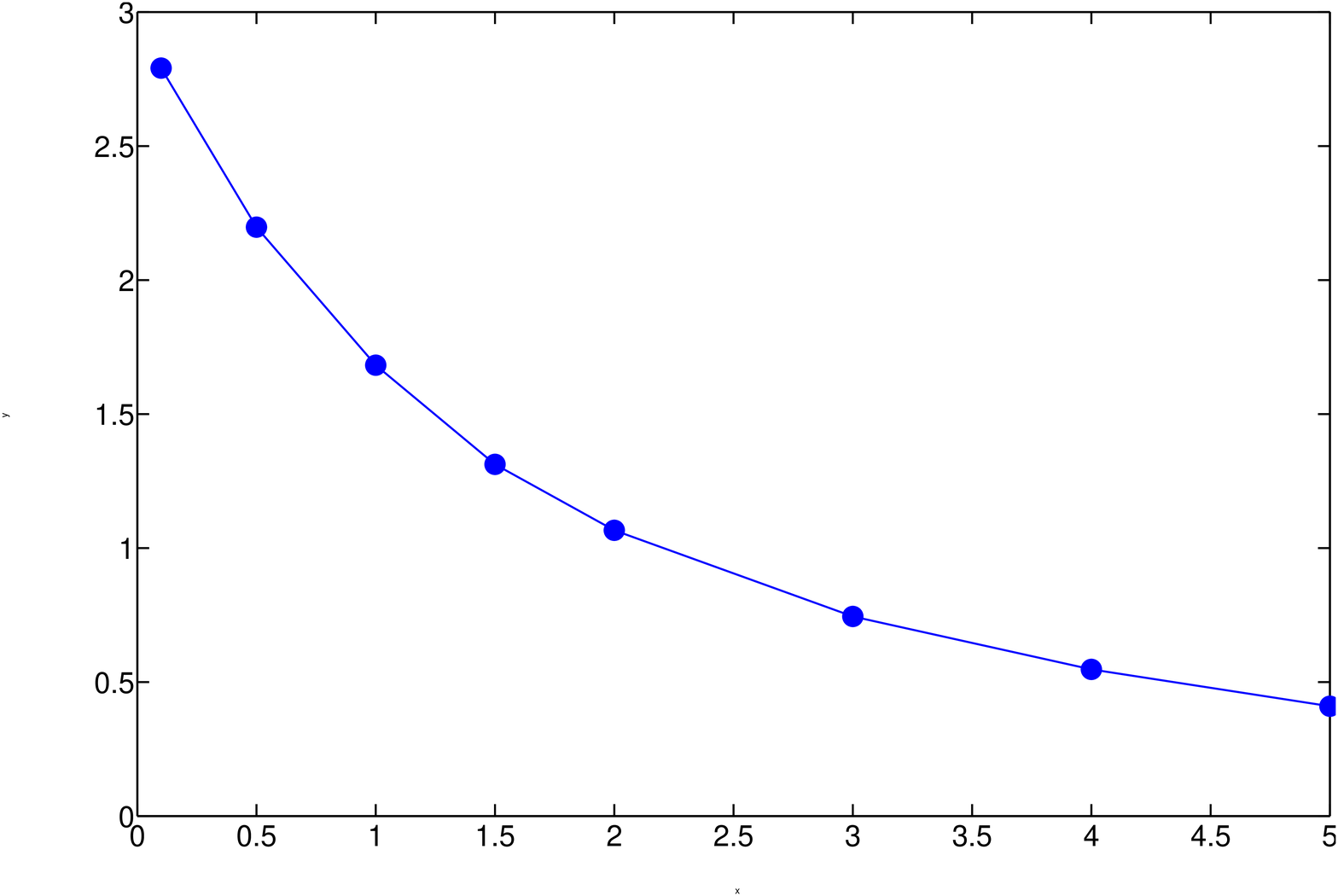}
\caption{(\textit{a}) Time-integrated p.d.f. of the largest eigenvalue of $\mathsfbi{M}$
for (from bottom to top) $\mathit{Ca}=0.21,0.32,0.51$ and different values of $\mu$.
The p.d.f.s corresponding to different values of $\mathit{Ca}$ are translated vertically;
(\textit{b}) exponential decay rate of the number of surviving drops as a function
of $\mu$ for $\mathit{Ca}=0.6$.
}
\label{fig:mu}
\end{figure}

The dependence of the deformation and breakup statistics on $\mu$ is
shown in figure~\ref{fig:mu}. For small values of $\mathit{Ca}$,
the slope of $\mathscr{P}(m_1^2)$ varies with $\mu$ and
is steeper for larger viscosity ratios (figure~\ref{fig:mu}(\textit{a})). 
It saturates to $-1$ beyond
the critical capillary number, but the transition to the supercritical
regime is slower for larger values of $\mu$. 
These results differ somewhat from those of \citet*{BMV14}. The
discrepancy may be explained by considering the time scales associated
with the breakup process. Whereas the time-integrated
statistics displays a weak dependence on the viscosity ratio, 
the time scale over which breakup occurs depends strongly on $\mu$,
and the breakup process considerably slows down as $\mu$ increases
(figure~\ref{fig:mu}(\textit{b})). For large values of $\mu$, very long Lagrangian trajectories
therefore need to be considered in order to compute
$\mathscr{P}(m_1^2)$; otherwise small deformations are privileged 
and the slope of $\mathscr{P}(m_1^2)$ may be steeper than it actually should be.
This point is elucidated further in \S~\ref{sect:analytical}.

\begin{figure}
\centering
\psfrag{y}[c]{$[\overline{T}(1)-\overline{T}(\rho_0)]/\tau_\eta$}
\psfrag{x}[c]{$\rho_0$}
\psfrag{a}[l]{$\rho_0^{\gamma}$}
\includegraphics[width=0.478\textwidth]{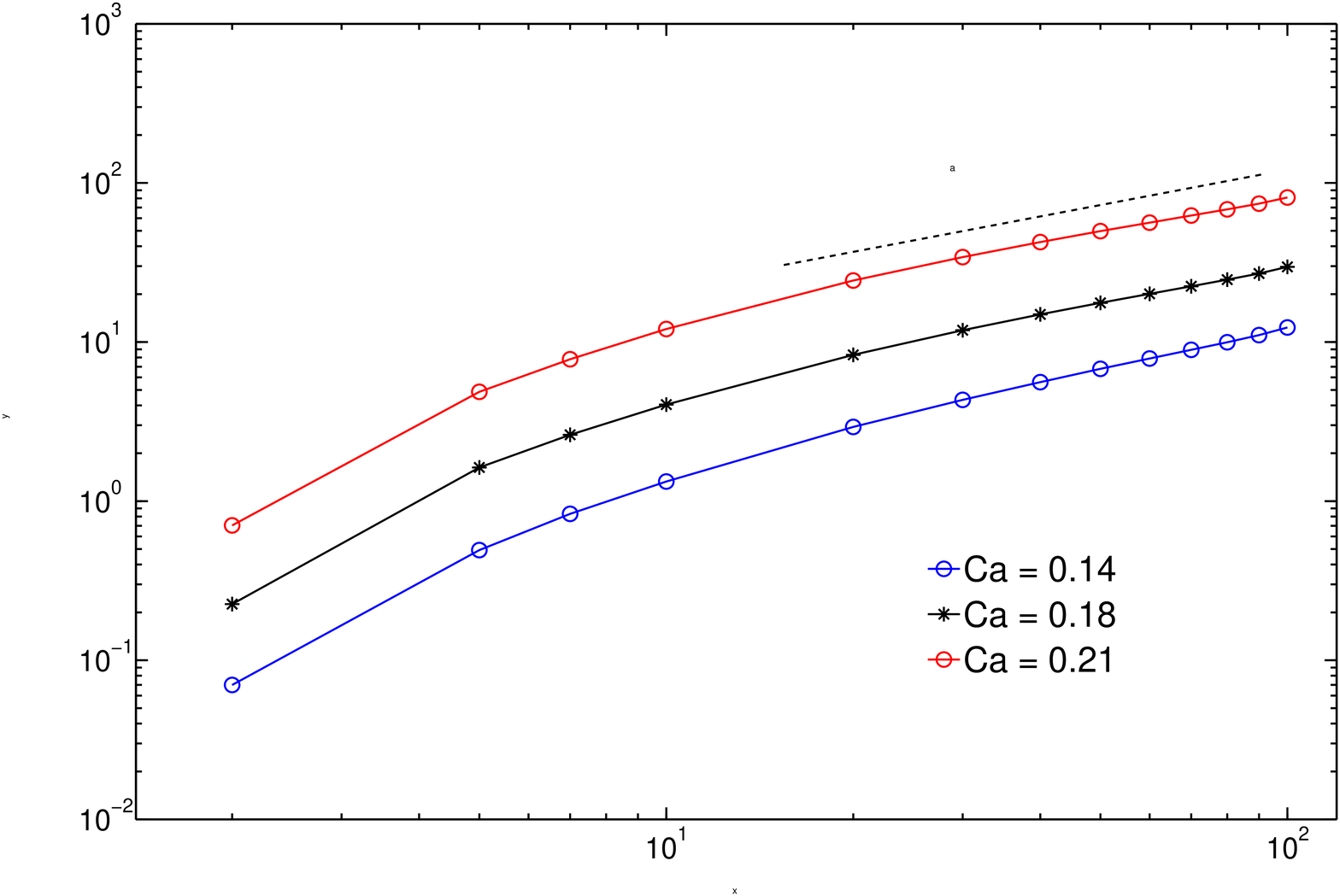}%
\psfrag{a}[l]{$\rho_0^{-\gamma}$}%
\psfrag{y}[c]{$\overline{T}(\rho_0)/\tau_\eta$}%
\hfill%
\includegraphics[width=0.514\textwidth]{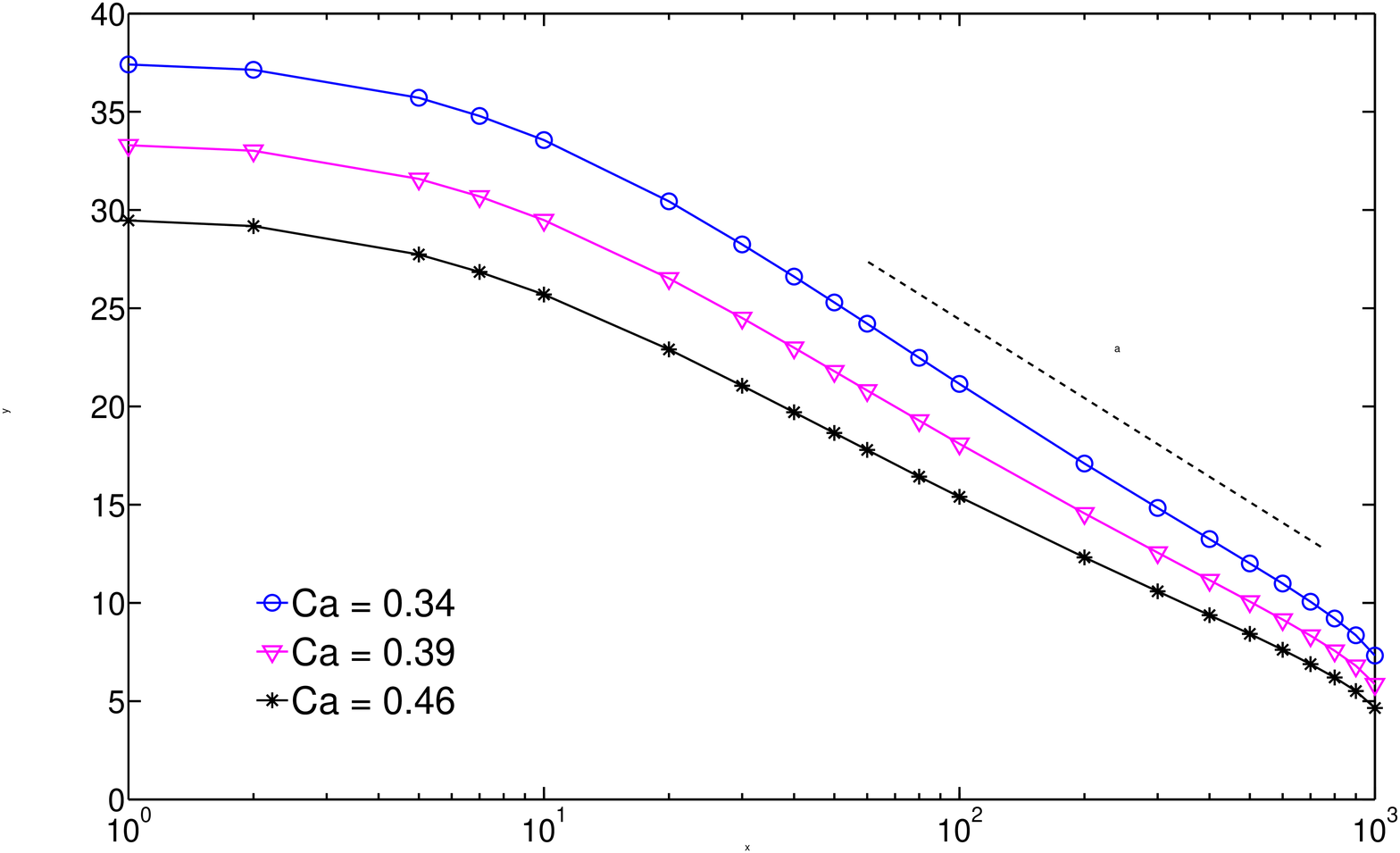}
\caption{Mean lifetime as a function of the initial aspect ratio for
(\textit{a}) $\mu=1$, $\alpha=10^2$, $\mathit{Ca}<\mathit{Ca}_c$ and
(\textit{b}) $\mu=1$, $\alpha=10^3$, $\mathit{Ca}>\mathit{Ca}_c$.
In (\textit{a}) the $\mathit{Ca}=0.206$ and $\mathit{Ca}=0.137$ curves
are multiplied by a factor 3 and 1/3, respectively, in order to
make the three curves distinguishable.
}
\label{fig:exit-time}
\end{figure}

Finally, we consider the mean lifetime of a drop, $\overline{T}(\rho_0)$, i.e. the mean 
time it takes for a drop of initial aspect ratio $\rho_0$ to break.
Two different behaviours are found depending on whether $\mathit{Ca}$ exceeds or not its
critical value. The mean lifetime $\overline{T}(\rho_0)$ decreases as
a power law of $\rho_0$ if $\mathit{Ca}<\mathit{Ca}_c$ and logarithmically
if $\mathit{Ca}>\mathit{Ca}_c$ (figure~\ref{fig:exit-time}).

The deformation and breakup statistics presented above is derived analytically in the next Section.

\section{Analytical predictions}
\label{sect:analytical}

For large deformations, the model of \citet{MM98} is statistically equivalent
to a vector model proposed by \citet*{ORL82}.
The assumptions on the drop and on the external fluid 
are essentially the same, and the 
semi-major axis $\bm r$ of the drop satisfies the equation:
\begin{equation}
\label{eq:r}
\dot{\bm r}=\mathsfbi{G}\,\bm r
-\dfrac{f_1(\mu)}{2\tau}\,\bm r
+\sqrt{\frac{\req^2 f_1(\mu)}{\tau}}\,\bm\xi(t),
\end{equation}
where $\req$ is the drop equilibrium size and
$\bm\xi(t)$ is white noise describing thermal fluctuations.
Although thermal noise does not appear in the original model of \citet*{ORL82}, 
it is included
in \eqref{eq:r} in order to regularize the p.d.f. of $\bm r$ at $r=0$. It
is anyway expected to play a minor role 
when the flow is turbulent or when
deformations larger than $\req$ are considered.
In this linear model, the condition for drop breakup is expressed in terms
of $r=\vert\bm r\vert$, i.e. it is assumed that a drop breaks if $r$ exceeds
a threshold size $\ell$.

Equation~\eqref{eq:r} is closely related to \eqref{eq:M}.
Indeed, from the vector $\bm r$ one can form the second-rank tensor
$\mathsfbi{M}=\langle\bm r\otimes\bm r\rangle_\xi$
($\langle\cdot\rangle_\xi$ denotes the average over the realizations 
of $\bm\xi(t)$),
which evolves according to the equation \citep*{ORL82}:
\begin{equation}
\label{eq:rr}
\dot{\mathsfbi{M}}=\mathsfbi{G}\mathsfbi{M}+\mathsfbi{M}\mathsfbi{G}^\top
-\dfrac{f_1(\mu)}{\tau}[\mathsfbi{M}-\req^2\mathsfbi{I}].
\end{equation}
The only difference between \eqref{eq:M} and \eqref{eq:rr}
is in the coefficient of the identity, which in \eqref{eq:M} 
preserves the volume of the drop whereas
does not enjoy this property in \eqref{eq:rr}.
Notwithstanding, 
this term is negliglible in both models
when the drops are sufficiently deformed.
Moreover, when $r$ is large $\mathsfbi{M}\approx\bm r\otimes\bm r$, so
$r^2$ is the largest eigenvalue of $\mathsfbi{M}$ and $\bm r$ the associate
eigenvector.
The statistics of large drop deformations can therefore be deduced
from \eqref{eq:r}, and potential discrepancies between the two
approaches are only expected for small deformations 
\citep[see][for a more detailed discussion of this point in the $\mu=1$ case]{VPBT15}.


Let us introduce the Kubo number $\mathit{Ku}=\lambda\tau_c$, 
where $\tau_c$ is the correlation time of $\bnabla\bm u$.
In three-dimensional homogeneous and isotropic turbulence
$\mathit{Ku}\approx 0.6$ \citep{GP90,BBBCLMT06,WG10}.
However, for $\mu=1$, it was shown in \citet{MV11} that
as long as $\mathit{Ku}\lesssim 1$,
the p.d.f. of $r$ does not depend  upon 
$\mathit{Ku}$ appreciably.
Furthermore, in \citet*{BMV14} the qualitative features of the
statistics of drop deformation were found not to depend significantly
on the intermittency of the turbulent flow.
To make analytical progress, we therefore study
\eqref{eq:r} under the assumption
that $\bnabla\bm u$ has a Gaussian statistics and $\tau_c$ vanishes.
More specifically, we assume that $\mathsfbi{\Omega}$
and $\mathsfbi{S}$ are zero-mean Gaussian processes with
correlations: $\langle{\Omega}_{ij}(t){\Omega}_{pq}(t')\rangle=
(d+2)C(\delta_{ip}\delta_{jq}-\delta_{jp}\delta_{iq})\delta(t-t')$ and
$\langle{S}_{ij}(t){S}_{pq}(t')\rangle=
dC(\delta_{ip}\delta_{jq}+\delta_{iq}\delta_{jp}
-2\delta_{ij}\delta_{pq}/d)\delta(t-t')$, where $d$ is the spatial 
dimension of the flow and 
$C>0$ determines the amplitude of the fluctuations of $\bnabla\bm u$. 
In this setting, $\mathsfbi{G}(t)$ is a multiplicative noise
and is interpreted in the Stratonovich sense \citep{FGV01}.
The form of the correlations ensures that the flow is incompressible and
statistically isotropic \citep[e.g.][]{BK97}.
In addition, the Lyapunov exponent of this flow is $\lambda=Cd(d-1)$ \citep{LJ84,LJ85}.

Owing to statistical isotropy, at long times
the p.d.f. of $r$, $p(r,t)$, satisfies the Fokker--Planck equation:
\begin{equation}
\label{eq:FPE}
\partial_t p = -\partial_r(D_1 p)+\partial_r^2(D_2 p),
\end{equation}
where time has been rescaled by $2\tau/f_1(\mu)$ (with a slight abuse of notation 
we continue to denote the rescaled time by $t$) and
\begin{equation}
\label{eq:coefficients}
D_1(r)=\left[\dfrac{2(d+1)\gamma(\mu)\mathit{Ca}}{d}
-1\right]r+(d-1)\dfrac{\req^2}{r},
\quad 
D_2(r)=\dfrac{2\gamma(\mu)\mathit{Ca}}{d}\,r^2+\req^2
\end{equation}
with $\gamma(\mu)=f_2(\mu)/f_1(\mu)$.
This equation can be obtained from the $\mu=1$ case
\citep[see][]{CMV05} by noting that the vorticity tensor does not 
contribute to the time evolution of $p(r,t)$.
The assumptions that $r$ is a positive quantity and drops break at $r=\ell$ are
implemented by imposing
a reflecting boundary condition at $r=0$ ($D_1 p-\partial_r(D_2 p)=0$
at $r=0$) and an absorbing boundary condition at $r=\ell$ 
($p(\ell,t)=0$), respectively.

The form of the coefficients $D_1(r)$ and $D_2(r)$ shows that changing the
viscosity ratio merely rescales $\mathit{Ca}$ by a factor of $\gamma(\mu)$.
Also note that $\gamma(\mu)$ depends weakly upon $\mu$, since it varies
from $\gamma(0)=2$ to $\gamma(\infty)=19/8=2.375$.

\subsection{Time-integrated distribution of drop sizes}

Equation~\eqref{eq:FPE} can be used to derive the 
power-law behaviour of the time-integrated p.d.f $\mathscr{P}(r)$
as well as its dependence on the initial drop-size distribution.
From \eqref{eq:FPE}, $\mathscr{P}(r)$ satisfies:
\begin{equation}
\label{eq:time-integrated}
-\dfrac{\rd}{\rd r}(D_1\mathscr{P})+
\dfrac{\rd^2}{\rd r^2}(D_2\mathscr{P})=
-p(r,0)
\end{equation}
with boundary conditions: 
$D_1\mathscr{P}-\partial_r(D_2\mathscr{P})=0$ at $r=0$
and $\mathscr{P}(\ell)=0$.
To obtain \eqref{eq:time-integrated},
we have used the fact that in the presence of an absorbing boundary
$\lim_{t\to\infty}p(r,t)=0$ for all $r$.
It is now assumed that $p(r,0)=\delta(r-r_0)$ with $r_0>\req$,
i.e. a monodisperse initial distribution.
Integrating \eqref{eq:time-integrated} from 0 to $r$ and using
the reflecting boundary condition at $r=0$ yields:
\begin{subnumcases}{\label{eq:complete}
-D_1(r)\mathscr{P}(r)+\dfrac{\rd}{\rd r}[D_2(r)\mathscr{P}(r)]=}
\label{eq:1st-order-a}
0 & \text{if $0\leqslant r<r_0$},
\\
\label{eq:1st-order-b}
-1 & \text{if $r_0<r\leqslant\ell$}.
\end{subnumcases}
The solution of \eqref{eq:complete} takes the form~\citep{R89}:
\begin{equation}
\label{eq:Pcal}
\mathscr{P}(r)\propto
\begin{cases}
\mathrm{e}^{-\Phi(r)}[\varphi(\ell)-\varphi(r_0)] & \text{if $0\leqslant r\leqslant r_0$},
\\
\mathrm{e}^{-\Phi(r)}[\varphi(\ell)-\varphi(r)] & \text{if $r_0<r\leqslant \ell$}
\end{cases}
\end{equation}
with
\begin{equation}
\label{eq:definitions}
\Phi(r)=\ln D_2(r)-\int_{r_1}^r\frac{D_1(\zeta)}{D_2(\zeta)}\,
\mathrm{d}\zeta,
\qquad
\varphi(r)=\int_{r_1}^r \dfrac{\mathrm{e}^{\Phi(\zeta)}}{D_2(\zeta)}\,\mathrm{d}\zeta,
\end{equation}
where the specific value of $r_1$ is irrelevant.
To examine the behaviour of $\mathscr{P}(r)$ for $\req\ll r\ll\ell$,
we now insert the limiting forms of $D_1(r)$ and $D_2(r)$ for $\req\to 0$
into \eqref{eq:definitions}
and obtain $\mathrm{e}^{\Phi(r)}\sim r^{\beta}$ 
and $\varphi(r)\sim r^{\beta-1}$ with $\beta=1-d+d/2\gamma(\mu)\mathit{Ca}$.
Therefore, there exists a critical value of the 
capillary number, $\mathit{Ca}_c=1/2\gamma(\mu)$, such that for
$\mathit{Ca}<\mathit{Ca}_c$
\begin{subnumcases}{\label{eq:pdf-1}\mathscr{P}(r)\sim}
\big(\ell^{\beta-1}-r_0^{\beta-1}\big)
r^{-\beta} & \text{if $\req\ll r \ll r_0$}\\
\label{eq:pdf-subcrit}
\ell^{\beta-1}r^{-\beta} & \text{if $r_0\ll r \ll \ell$},
\end{subnumcases}
whereas for $\mathit{Ca}>\mathit{Ca}_c$
\begin{subnumcases}{\label{eq:pdf-2}\mathscr{P}(r)\sim}
\big(\ell^{\beta-1}-r_0^{\beta-1}\big)
r^{-\beta} & \text{if $\req\ll r \ll r_0$}\\
\label{eq:pdf-supercrit}
r^{-1} & \text{if $r_0\ll r \ll \ell$}.
\end{subnumcases}
(The exact form of $\mathscr{P}(r)$ over 
the entire interval $0\leqslant r\leqslant 
\ell$ may be calculated by using the full expressions of $D_1(r)$ and
$D_2(r)$ in \eqref{eq:coefficients} and involves a hypergeometric function;
the details, however, are omitted.)
Since $\gamma(\mu)$ depends weakly on $\mu$, the same holds true for
$\mathit{Ca}_c$.
The above value of $\mathit{Ca}_c$ was also found by \citet*{BMV14}
for more general flows; they applied a criterion based on the statistics
of the finite-time Lyapunov exponents of the flow that was previously used to
study the deformation of flexible polymers \citep{BFL01}.
Likewise, the prediction of \citet*{BMV14} for the exponent $\beta$ in
the $\mathit{Ca}<\mathit{Ca}_c$ case
reduces to the expression above when $\bnabla\bm u$ has the properties considered here.

The scaling of $\mathscr{P}(r^2)$ can be deduced from that of
$\mathscr{P}(r)$ by using: $\mathscr{P}(r^2)=
\frac{1}{2}r^{-1}\mathscr{P}(r)$.
The above power-law behaviours thus reproduce the numerical results
shown in figures~\ref{fig:m_1} and~\ref{fig:initial_condition}
for the time-integrated p.d.f.s of the 
squared length of the semi-major axis.
It should be noted that whereas the power-law behaviour of $\mathscr{P}(r)$
for small $r$ is
specific to a monodisperse initial distribution,
the large-$r$ power law 
holds for any $p(r,0)$ that vanishes beyond a given $r_\star<\ell$.
Integrating \eqref{eq:time-integrated} from 0 to $r>r_\star$ indeed yields
\eqref{eq:1st-order-b} and hence \eqref{eq:pdf-subcrit} or
\eqref{eq:pdf-supercrit} depending on the value of $\mathit{Ca}$.
If, by contrast, the initial size of drops can approach $\ell$, in general
$\mathscr{P}(r)$ does not display a power-law behaviour.

\subsection{Time-dependent distribution of drop sizes and breakup frequency}

\begin{figure}
\centering
\includegraphics[width=0.5\textwidth]{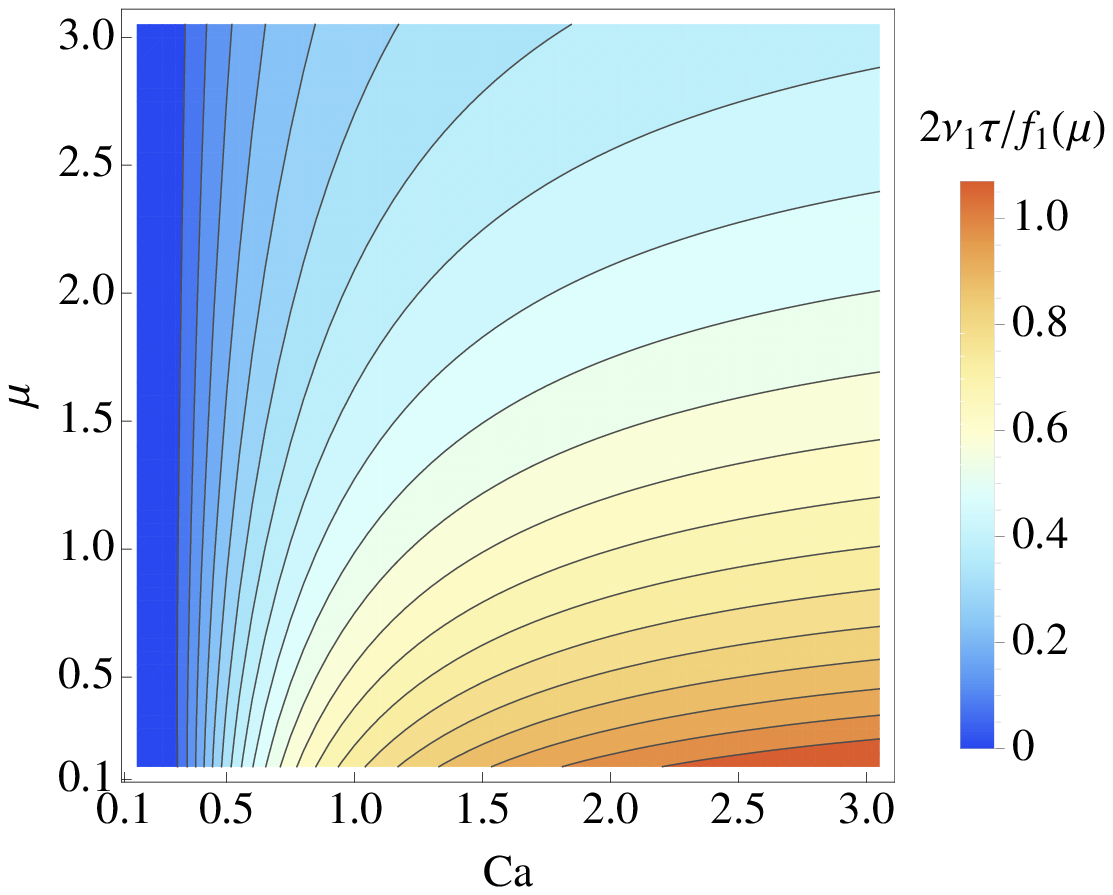}%
\hfill%
\includegraphics[width=0.41\textwidth]{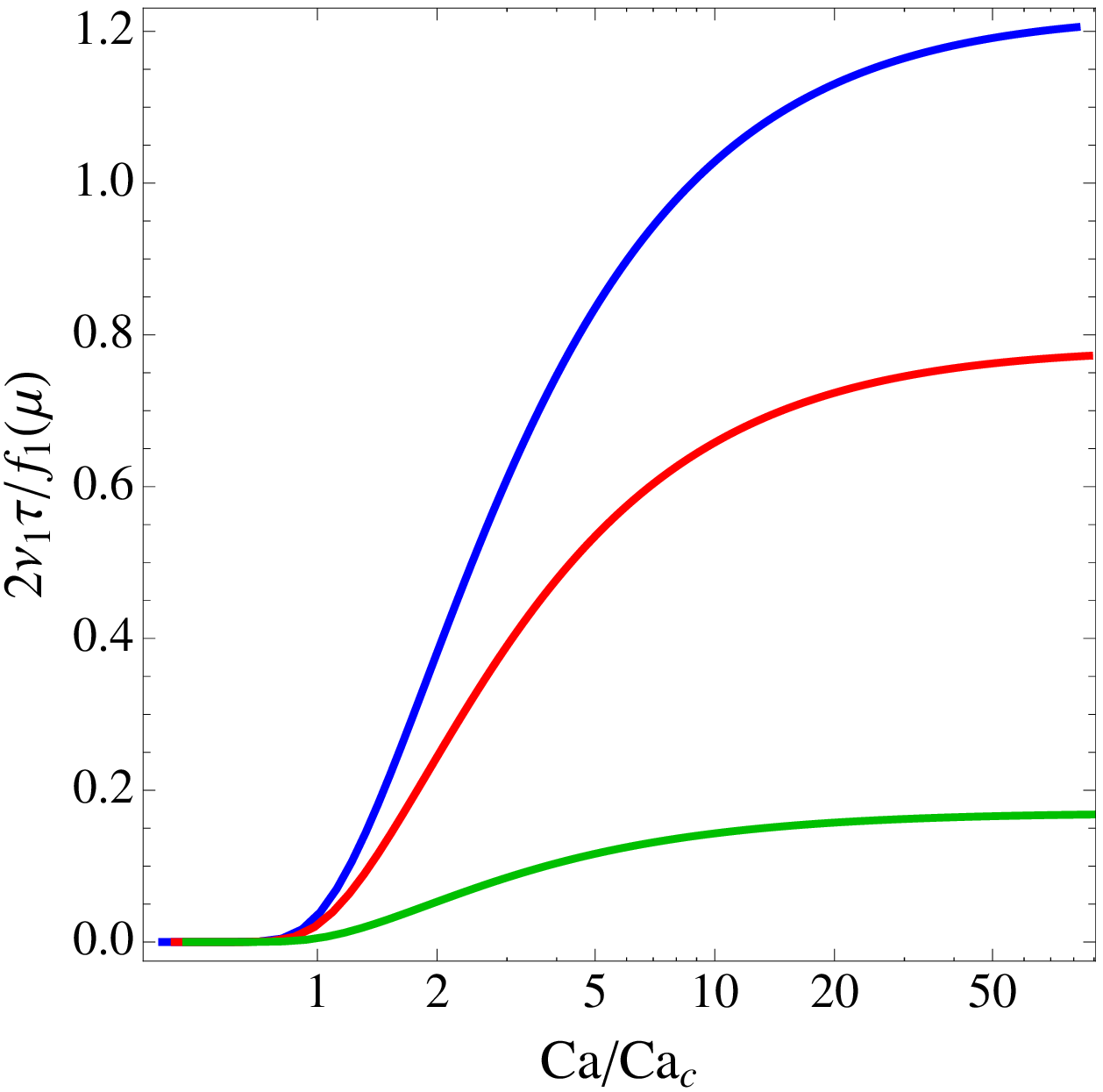}
\caption{Exponential decay rate of the number of drops
for $d=3$ and $\ell=10^3$ (\textit{a}) as a function
of $\mu$ and $\mathit{Ca}$ and
(\textit{b}) as a function of the capillary number rescaled
by its critical value for fixed $\mu=0.1,1,10$ (from top to bottom).}
\label{fig:decay-rate}
\end{figure}

The eigenfunctions of the Fokker--Planck operator 
that satisfy the reflecting boundary condition at $r=0$ are 
of the form 
$f_\nu(r)=r^{d-1} {_2F_1}(c^+_\nu,c_\nu^-,d/2,-\epsilon r^2)$ \citep*{CMV05}, 
where $_2F_1$ is the Gauss hypergeometric function with
$\epsilon=2\gamma(\mu)\mathit{Ca}/d\req^2$ and 
\[
c^{\pm}_\nu=\dfrac{d}{4}\left[\dfrac{1}{2\gamma(\mu)\mathit{Ca}}+1\right]
\mp \dfrac{1}{4}
\sqrt{d^2\left[\dfrac{1}{2\gamma(\mu)\mathit{Ca}}-1\right]^2
-\dfrac{2 d \nu}{\gamma(\mu)\mathit{Ca}}
}.
\]
The absorbing boundary condition
$f_\nu(\ell)=0$ selects a discrete set of acceptable eigenfunctions.
The p.d.f. of $r$ can thus be expanded as
$p(r,t)=\sum_{n=1}^\infty a_n \mathrm{e}^{-\nu_n t}f_{\nu_n}(r)$,
and hence
\begin{equation}
\label{eq:asymptotic}
p(r,t)\sim\mathrm{e}^{-\nu_1 t}f_{\nu_1}(r) 
\quad \text{as $t\to\infty$}.
\end{equation}
This result confirms that at long times $p(r,t)$ approaches
an asymptotic shape, but this does not show a power-law behaviour
(figure~\ref{fig:time-dependence}(\textit{a})).
From \eqref{eq:asymptotic}, the fraction of drops surviving at time $t$
decays as
\begin{equation}
N(t)/N(0)\equiv \int_0^\ell p(r,t)\mathrm{d}r
\sim\mathrm{e}^{-\nu_1 t}  \quad \text{as $t\to\infty$},
\end{equation}
where $\nu_1$ is the smallest solution of the equation $f_{\nu_n}(\ell)=0$.
Figure \ref{fig:decay-rate}(\textit{a}) shows that
the decay rate of the drop number increases rapidly as a function
of $\mathit{Ca}$ when $\mathit{Ca}$ exceeds its critical value, whereas
it decreases as a function of $\mu$. In addition, although
$\mathit{Ca}_c$ depends weakly on $\mu$, the transition to the supecritical
regime is much steeper at small $\mu$
(see also figure~\ref{fig:decay-rate}(\textit{b})).
These results reproduce the behaviours observed in the numerical simulations
(figures~\ref{fig:time-dependence}(\textit{b)} and~\ref{fig:mu}).

\subsection{Mean lifetime of a drop}
The average time it takes for a drop of initial size $r_0$ to break
can be calculated from $\mathscr{P}(r)$ as follows.
Consider the transition probability $p(r,t\vert r_0,0)$, which is the
solution of \eqref{eq:FPE} that satisfies the
initial condition $p(r,0\vert r_0,0)=\delta(r-r_0)$.
Let $T(r_0)$ be the time it takes for the drop to break in a given realization
of the flow and of thermal noise, and let $\mathbb{P}(r_0,t)$ be
the probability of $T(r_0)$ taking the value $t$.
Note that $\mathbb{P}(r_0,t)=-\partial_t F$, where $F(r_0,t)=\int_t^\infty 
\mathbb{P}(r_0,s)\rd s$
is the probability that $T(r_0)\geqslant t$ and can be written as
$F(r_0,t)=\int_0^\ell p(r,t\vert r_0,0)\rd r$.
Therefore, the average of $T(r_0)$ is \citep{G83}:
\begin{equation}
\overline{T}(r_0)=
\int_0^\infty t\,\mathbb{P}(r_0,t)\rd t=
-\int_0^\infty t[\partial_t F(r_0,t)] \rd t=\int_0^\infty F(r_0,t)\,\rd t,
\end{equation}
where we used $\lim_{t\to\infty}F(t)=0$, a consequence of the absorbing
boundary condition for $p(r,t\vert r_0,0)$.
By changing the order of integration, we finally obtain:
\begin{equation}
\label{eq:average-time}
\overline{T}(r_0)=\int_0^\ell  \mathscr{P}(r) \rd r,
\end{equation} 
where $\mathscr{P}(r)$ is the solution of 
\eqref{eq:time-integrated} corresponding to the initial condition
$p(r,0)=\delta(r-r_0)$.
Inserting now the asymptotic behaviours 
\eqref{eq:pdf-subcrit} and \eqref{eq:pdf-supercrit}
into \eqref{eq:average-time} yields:
\begin{equation}
{\overline{T}(r_0)}\sim
\begin{cases}
\left(\dfrac{\ell}{\req}\right)^{\beta-1}-\left(\dfrac{r_0}{\req}\right)^{\beta-1}
& \text{if $\mathit{Ca}<\mathit{Ca}_c$},
\\[5mm]
\ln\bigg(\dfrac{\ell}{r_0}\bigg) & \text{if $\mathit{Ca}>\mathit{Ca}_c$},
\end{cases}
\label{eq:lifetime}
\end{equation}
as seen in figure~\ref{fig:exit-time}.

\section{Improved version of the model of \cite{MM98}}
\label{sect:improved-f2}

\begin{figure}
\includegraphics[width=0.415\textwidth]{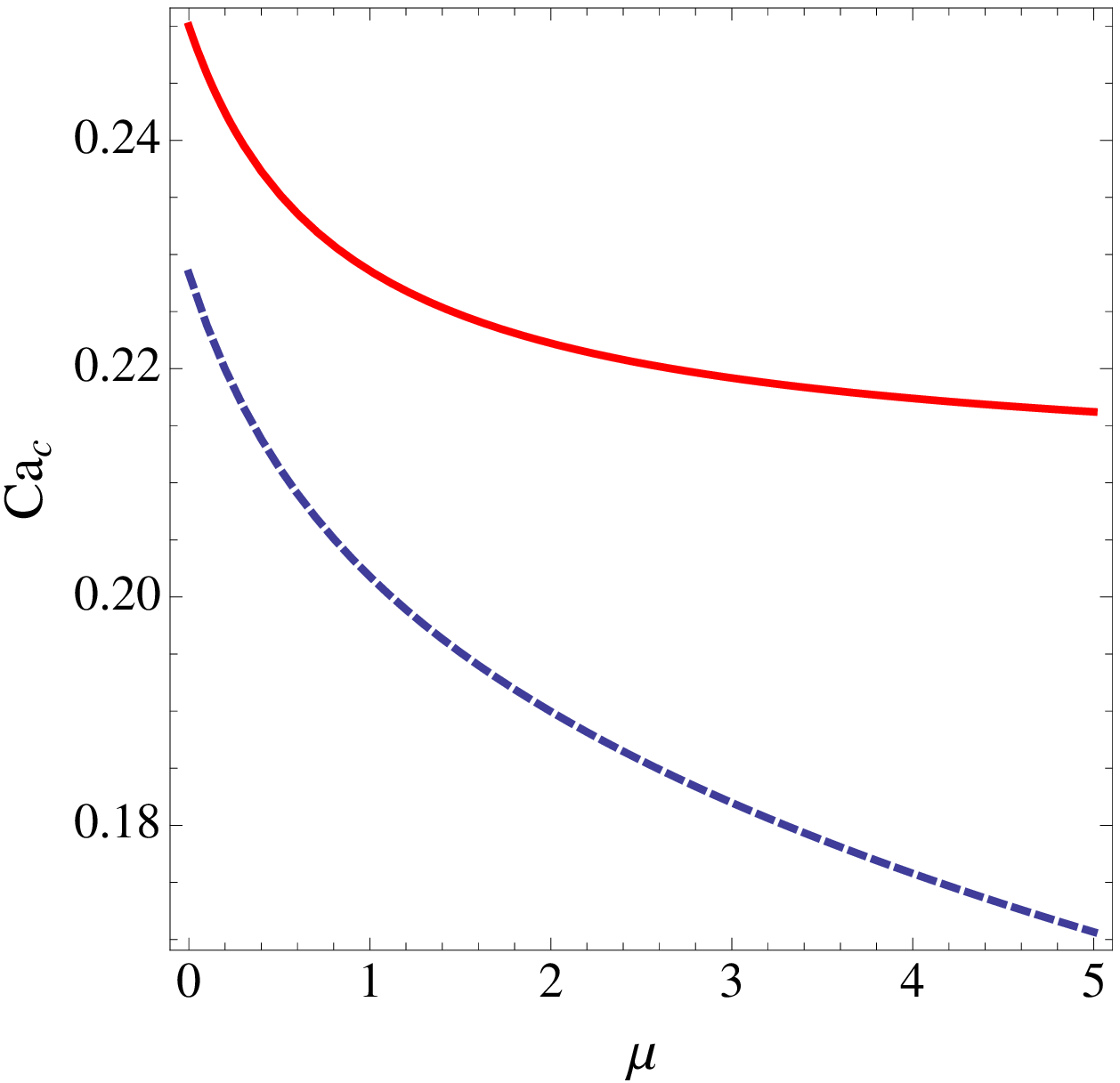}%
\hfill%
\includegraphics[width=0.5\textwidth]{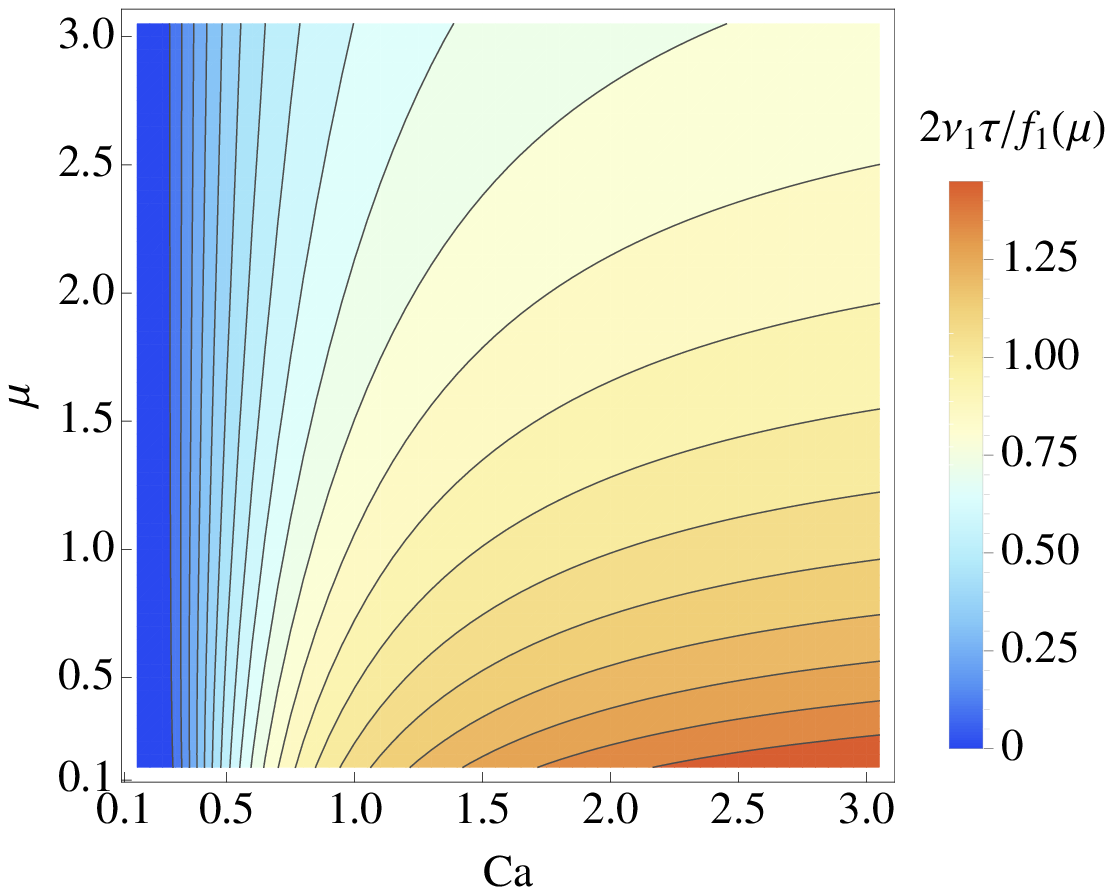}%
\caption{(\textit{a}) Dependence of the critical capillary number
on the viscosity ratio in the original model (red solid curve) 
and in the improved model (blue dashed curve);
(\textit{b}) Exponential decay rate of the number of drops
for $d=3$ and $\ell=10^3$ as a function of $\mathit{Ca}$ and $\mu$
for the improved model.
}
\label{fig:comparison-Ca_c}
\end{figure}

\cite{MM98} proposed a modification of their model that improves
the agreement with experimental data
for high viscosity ratios and large capillary numbers.
In the modified model,
the coefficient in front of the strain tensor also depends
on $\mathit{Ca}$, i.e., $f_2(\mu)$ is replaced with
\begin{equation}
\tilde{f}_2(\mu,\mathit{Ca})=f_2(\mu)+\dfrac{3(\sigma\mathit{Ca})^2}%
{2+6(\sigma\mathit{Ca})^2},
\end{equation}
where the coefficient $\sigma$ accounts for the fact that here
$\mathit{Ca}$ is defined
in terms of $\lambda$ (hence in our case $\sigma=1.72$).
The above expression still reproduces the theoretical limits,
for both small $\mathit{Ca}$ and large $\mu$, as well as
the affine deformation of the drop
when $\mu=1$ and $\mathit{Ca}\to\infty$.
Using $\tilde{f}_2$ instead of $f_2$ yields significantly more accurate
predictions for a pure shear; 
for an elongational flow, the effect is much weaker \citep{MM98}.

It should be noted that the original model and the 
improved one can be mapped into each other by suitably modifying
the viscosity ratio and the capillary relaxation time. 
The original model with parameters $\mu$, $\tau$
is indeed the same as the improved one with parameters $\mu'$, $\tau'$,
where $\mu'$ and $\tau'$ are the solutions of the system:
\begin{equation}
f_1(\mu')/\tau'=f_1(\mu)/\tau,
\qquad
\tilde{f}_2(\mu',\sigma\lambda\tau')=f_2(\mu).
\end{equation}
Therefore, for fixed $\mu$ and $\mathit{Ca}$, the results described
in the previous Sections also hold for the improved model,
provided the parameters are suitably adjusted.
The reader should note that such a nonlinear transformation 
of the parameters leads to a slight variation, quantitatively, of the results
without changing the overall picture.
It is nonetheless important to examine the effect of the modified
coefficient $\tilde{f}_2$ on quantities such as
the critical capillary number, the rate of decay of the
drop fraction, and the exponent $\beta$
that defines the power-law behaviour of both the p.d.f. of the size and 
the mean lifetime.
This is achieved by replacing $\gamma(\mu)$ in 
\S~\ref{sect:analytical} with 
$\tilde{\gamma}(\mu,\mathit{Ca})=\tilde{f}_2(\mu,\mathit{Ca})/f_1(\mu)$.
Thus, the differences between the two versions of the model are mainly due
to the fact that $\gamma(\mu)$ varies weakly with $\mu$ and is bounded
for $\mu\to\infty$, whereas
$\tilde{\gamma}(\mu,\mathit{Ca})\to\infty$ in the same limit.
It is shown below that, for a turbulent flow,
these differences impact our predictions only marginally.

When the improved model is considered,
the critical capillary number is the solution of the cubic
equation $\tilde{\gamma}(\mu,\mathit{Ca}_c)\mathit{Ca}_c=1/2$.
It can be checked that the discriminant of this equation is negative
for all values of $\mu$ and hence there is only one real root.
Figure~\ref{fig:comparison-Ca_c}(\textit{a}) 
compares the critical capillary number in
the original model and in the improved one.
In both cases, $\mathit{Ca}_c$ is a decreasing function of $\mu$.
The main difference is that, for $\mu\to\infty$,
$\mathit{Ca}_c$ tends to the asymptotic
value $4/19\approx 0.21$ in the original model, 
whereas it tends to zero in the improved one.
Nevertheless, for the values of $\mu$ typically found in experiments,
$\mathit{Ca}_c$ does not differ considerably in the two models.

\begin{figure}
\includegraphics[width=0.5\textwidth]{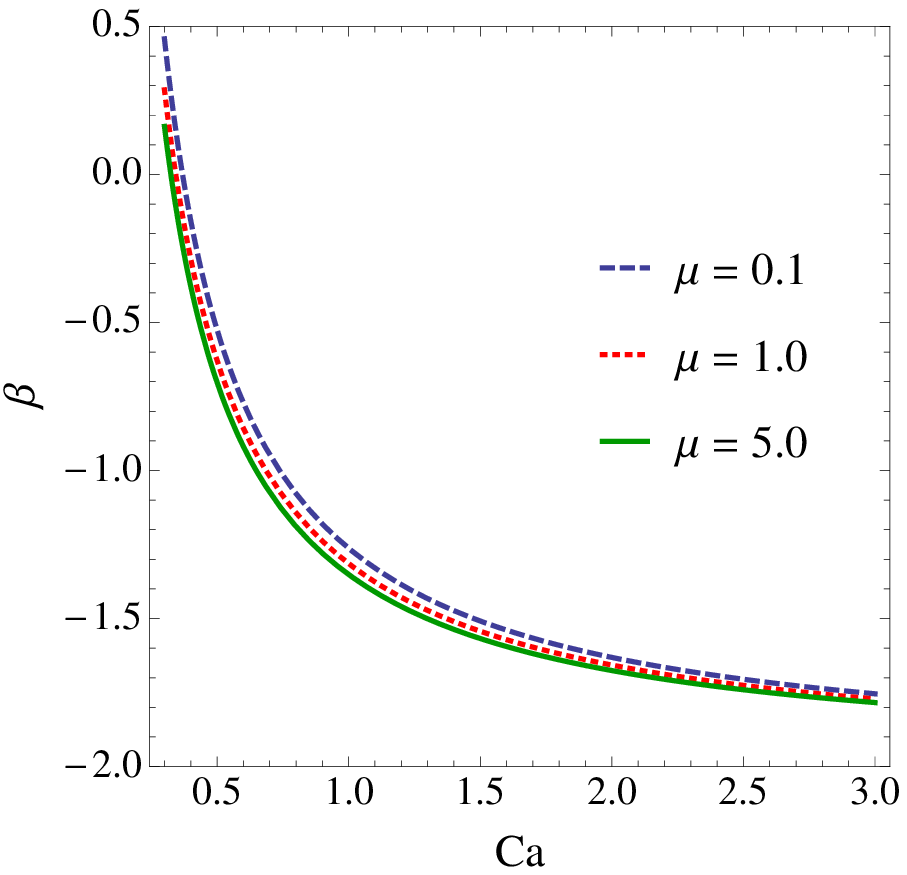}%
\hfill%
\includegraphics[width=0.5\textwidth]{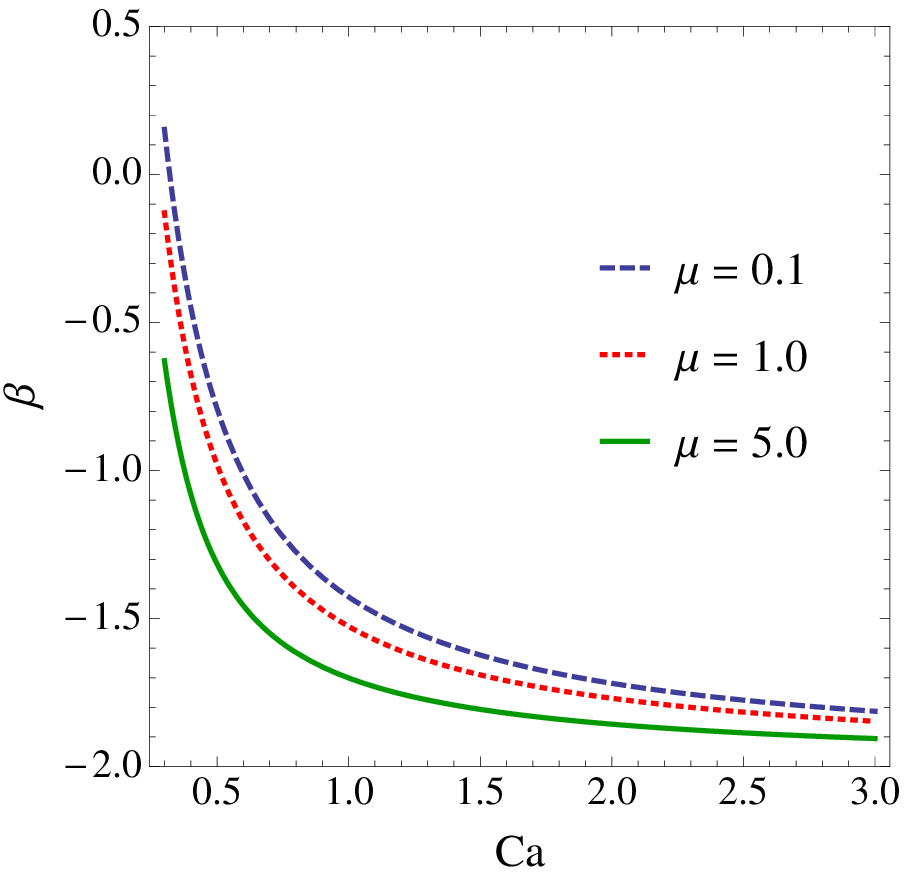}
\caption{ Exponent $\beta$ as a function of the capillary number
for $d=3$ (\textit{a}) in the original model and 
(\textit{b}) in the improved model.
}
\label{fig:beta}
\end{figure}

In the improved model, the rate of decay of the number of drops is slightly
greater and decreases less rapidly as a function on $\mu$,
(compare figures~\ref{fig:decay-rate}(\textit{a})
and~\ref{fig:comparison-Ca_c}(\textit{b})).
The exponent $\beta$ takes the form
$\beta=1-d+d/2\tilde{\gamma}(\mu,\mathit{Ca})\mathit{Ca}$;
it is smaller than in the original model
and varies more with $\mu$, as a consequence of the different behaviour
of $\tilde{\gamma}(\mu,\mathit{Ca})$ (figure~\ref{fig:beta}).
Both for $\beta$ and the decay rate,
the discrepancies between the two models are, however, small.

In conclusion, despite some quantitative differences, for realistic values of
$\mu$ and \textit{Ca} the qualitative
behaviour of the model of \cite{MM98} in a turbulent flow
is largely insensitive to the use of either $f_2(\mu)$ or $\tilde{f}_2(\mu,\mathit{Ca})$.

\section{Conclusions}

The Lagrangian dynamics of a sub-Kolmogorov drop in a turbulent flow is
determined by the statistics of the velocity gradient. 
Strong fluctuations of the strain along the 
trajectory of the drop highly modify the shape and the size of the drop and
ultimately break it. We have performed a detailed numerical and analytical
study of the deformation and breakup statistics of neutrally buoyant,
sub-Kolmogorov, ellipsoidal drops in homogeneous and isotropic turbulence
as a function of the capillary number, 
the viscosity ratio between the inner and outer fluids
and the initial drop-size distribution.
In particular, we have analytically derived some of the numerical
observations reported in \cite*{BMV14} and have extended the prediction
for the critical capillary number to viscosity ratios different from unity.
We have also examined further properties of the breakup process, such as the 
temporal dependence of the number of drops and of the statistics
of the size, the role of the initial distribution of the drop sizes,
and the mean lifetime of a drop.

Our study relies on the model of \citet{MM98}.
Potential extensions concern the impact on the
deformation and breakup dynamics of effects that are not taken
into account by this model
These include
deviations from the ellipsoidal shape, nonlinear deformations near to breakup,
large density contrasts between the fluids inside and
outside the drop, or secondary breakups.
More refined models of drop dynamics have indeed been 
proposed in the literature \citep[e.g.][]{M10}.
However, such models generally depend in a highly nonlinear way
on the shape of the drop, and this
renders their analytical study very challenging.

It would also be interesting to understand possible intermittency effects for 
such sub-Kolmogorov scale droplets~\citep{BMV14} (and also studied for larger droplets~\citep{PBSST12}) 
and if there are analogues of transparency effects, seen 
in oscillatory, laminar flows~\citep{MSBT18} for droplets in fully developed turbulence.

Finally, \cite{MM98} observe that, 
for $\mu=1$, their model is closely related to the Oldroyd-B model for 
solutions of flexible polymers \citep{BHAC87}. Likewise, when $\mu=1$
and hence $\mathsfbi{G}=\bnabla\bm u$
the vector model of \cite*{ORL82} reduces to the Hookean dumbbell model,
which describes the evolution of the 
end-to-end separation vector of a flexible polymer molecule 
in the limit in which nonlinear elastic effects are negligible \citep{BHAC87}.
Therefore, after appropriate redefinition of the parameters,
our results also apply to the degradation of polymers in turbulent flows.

\acknowledgments
The authors would like to thank A. Gupta and P. Perlekar
for fruitful discussions.
They also acknowledge the support of the Indo--French Centre for Applied
Mathematics (IFCAM)
and the EU COST Action MP 1305 `Flowing Matter'.
SSR acknowledges the support of the DAE and the DST (India) project ECR/2015/000361 and the 
F\'ed\'eration Doeblin.
Our simulations were performed on the cluster {\it Mowgli}
and the work station {\it Goopy} at the ICTS-TIFR.
D. V. acknowledges the hospitality of the Max Planck Institute for
Dynamics and Self-Organisation, where part of this work was done.

\appendix

\section{Cholesky decomposition of the tensor $\mathsfbi{M}$}

The Cholesky decomposition of $\mathsfbi{M}$ is $\mathsfbi{M}=\mathsfbi{L}\mathsfbi{L}^\top$, where 
$\mathsfbi{L}$ is a lower triangular matrix. The elements of $\mathsfbi{L}$ satisfy:
\begin{eqnarray*}
\dot{L}_{11}&=&
G_{11}L_{11}+G_{12}L_{21}+G_{13}L_{31}+c_\tau\left[
\dfrac{c_g}{L_{11}}-L_{11}\right]
\\[2mm]
\dot{L}_{21}&=&G_{21}L_{11}+G_{22}L_{21}+G_{23}L_{31}+G_{12}
\dfrac{L_{22}^2}{L_{11}}+G_{13}\dfrac{L_{32}L_{22}}{L_{11}}
-c_\tau\left[L_{21}+c_g\dfrac{L_{21}}{L_{11}^2}\right]
\\[2mm]
\dot{L}_{31}&=&G_{31}L_{11}+G_{32}L_{21}+G_{33}L_{31}+G_{12}
\dfrac{L_{22}L_{32}}{L_{11}}+G_{13}\dfrac{L_{32}^2+L_{33}^2}{L_{11}}
-c_\tau\left[L_{31}+c_g\dfrac{L_{31}}{L_{11}^2}\right]
\\[2mm]
\dot{L}_{22}&=&G_{22}L_{22}+G_{23}L_{32}-G_{12}
\dfrac{L_{21}L_{22}}{L_{11}}-G_{13}\dfrac{L_{32}L_{21}}{L_{11}}
+c_\tau\left[-L_{22}+\dfrac{c_g}{L_{22}}+c_g\dfrac{L_{21}^2}{L_{11}^2L_{22}}\right]
\\[2mm]
\dot{L}_{32}&=&G_{32}L_{22}+G_{33}L_{32}-G_{12}
\dfrac{L_{22}L_{31}}{L_{11}}-G_{13}\dfrac{L_{31}L_{32}}{L_{11}}
+G_{23}\dfrac{L_{33}^2}{L_{22}}-G_{13}\dfrac{L_{21}L_{33}^2}{L_{11}L_{22}}
\\ &&
+2c_\tau c_g\dfrac{L_{21}L_{31}}{L_{11}^2L_{22}}
+c_\tau\left[-L_{32}-c_g\dfrac{L_{32}}{L_{22}^2}-c_g\dfrac{L_{21}^2L_{32}}{L_{11}^2L_{22}^2}
\right]
\\[2mm]
\dot{L}_{33}&=&
G_{33}L_{33}-G_{13}\dfrac{L_{31}L_{33}}{L_{11}}-G_{23}
\dfrac{L_{32}L_{33}}{L_{22}}+G_{13}\dfrac{L_{21}L_{32}L_{33}}{L_{11}L_{22}}
-2c_\tau c_g\dfrac{L_{21}L_{31}L_{32}}{L_{11}^2L_{22}L_{33}}
\\ &&
+c_\tau\left[-L_{33}+\dfrac{c_g}{L_{33}}+c_g\dfrac{L_{31}^2}{L_{11}^2L_{33}}
+c_g\dfrac{L_{32}^2}{L_{22}^2L_{33}}
+c_g\dfrac{L_{21}^2L_{32}^2}{L_{11}^2L_{22}^2L_{33}}\right]
\end{eqnarray*}
with $c_\tau=f_1(\mu)/2\tau$ and $c_g=g(\mathsfbi{L}\mathsfbi{L}^\top)$
(the functions $f_1$ and $g$ are defined after \eqref{eq:M}).
The above equations can be derived by adapting to \eqref{eq:M}
the equations obtained in
\citet{VC03} for a constitutive model of viscoelastic fluid
(see also \citet{PMP06}, where a misprint is corrected in the 
equation for $L_{32}$).
The positivity of $L_{ii}$, $i=1,2,3$, is enforced by evolving $\ln L_{ii}$
instead of $L_{ii}$ \citep{VC03}.

\end{document}